\documentclass[showpacs,aps,prd,amsmath,amsfonts,amssymb,eqsecnum,nofootinbib,
  floatfix,secnumarabic,preprintnumbers,superscriptaddress,longbibliography]{revtex4-1}
\usepackage[colorlinks=true]{hyperref}
\usepackage{epsfig,slashed}

\begin{document}
\newcommand{\be}{\begin{equation}}
\newcommand{\ee}{\end{equation}}
\newcommand\bpm{\begin{pmatrix}}
\newcommand\epm{\end{pmatrix}}
\def\bsp#1\esp{\begin{split}#1\end{split}}
\renewcommand{\d}{{\rm d}} 
\newcommand{\etc}{{\it etc.}}
\newcommand{\ie}{{\it i.e.}}
\newcommand{\eg}{{\it e.g.}}
\newcommand{\lag}{{\cal L}} 
\newcommand{\gmu}{\gamma^\mu}
\newcommand{\feynrules}{{\sc FeynRules}}
\newcommand{\madgraph}{{\sc MadGraph}}
\newcommand{\madevent}{{\sc MadEvent}}
\newcommand{\madanalysis}{{\sc MadAnalysis}}
\newcommand{\fewz}{{\sc Fewz}}
\newcommand{\hathor}{{\sc Hathor}}
\newcommand{\pythia}{{\sc Pythia}}
\newcommand{\tauola}{{\sc Tauola}}
\newcommand{\delphes}{{\sc Delphes}}
\newcommand{\fastjet}{{\sc FastJet}}
\newcommand{\met}{\slashed{E}_T }

\preprint{CERN-PH-TH/2013-277}

\title{Monotop phenomenology at the Large Hadron Collider}

\author{Jean-Laurent Agram}
\affiliation{Groupe  de Recherche de Physique des Hautes Energies (GRPHE),
    Universit\'e de Haute-Alsace, IUT Colmar, 34 rue du Grillenbreit BP 50568, 
    68008 Colmar Cedex, France}
\author{Jeremy Andrea}
\affiliation{Institut Pluridisciplinaire Hubert Curien/D\'epartement Recherches
 Subatomiques, Universit\'e de Strasbourg/CNRS-IN2P3, 23 Rue du Loess,
 F-67037 Strasbourg, France}
\author{Michael Buttignol}
\affiliation{Groupe  de Recherche de Physique des Hautes Energies (GRPHE),
    Universit\'e de Haute-Alsace, IUT Colmar, 34 rue du Grillenbreit BP 50568, 
    68008 Colmar Cedex, France}
\author{Eric Conte}
\affiliation{Groupe  de Recherche de Physique des Hautes Energies (GRPHE),
    Universit\'e de Haute-Alsace, IUT Colmar, 34 rue du Grillenbreit BP 50568, 
    68008 Colmar Cedex, France}
\author{Benjamin Fuks}
\affiliation{Institut Pluridisciplinaire Hubert Curien/D\'epartement Recherches
 Subatomiques, Universit\'e de Strasbourg/CNRS-IN2P3, 23 Rue du Loess,
 F-67037 Strasbourg, France}
\affiliation{Theory Division, Physics Department, CERN, CH-1211 Geneva 23, Switzerland}

\date{\today}
\begin{abstract}
We investigate new physics scenarios where systems comprised of a single top quark
accompanied by missing transverse energy, dubbed monotops,
can be produced at the LHC. Following a simplified model approach,
we describe all possible monotop production modes via an effective theory
and estimate the sensitivity of the LHC, assuming 20~fb$^{-1}$ of collisions at
a center-of-mass energy of 8~TeV, to the observation of a monotop state.
Considering both leptonic and hadronic top quark decays,
we show that large fractions of the parameter space are reachable
and that new physics particles with masses ranging up to 1.5~TeV can leave hints
within the 2012 LHC dataset, assuming moderate new physics coupling strengths.
\end{abstract}


\maketitle

\section{Introduction}
After almost half a century, the Standard Model of particle physics
has been proved to largely describe all experimental
high-energy physics data. Despite its success, it however leaves
many questions unanswered so that it is believed to be the low-energy limit
of a more fundamental theory that is still to be discovered. Whilst there is currently no
hint for any physics beyond the Standard Model, new phenomena are still actively searched for
experimentally whereas on the theory side, model building activities are ongoing quite intensely
for the last decades.
In particular, the sector of the top quark whose mass is close to the electroweak
symmetry breaking scale is expected to be one of the key candle for the possible discovery
of new physics of any kind.

In the framework of the Standard Model, top quarks are mainly produced,
at the Large Hadron Collider (LHC), either in pairs
or singly via weak subprocesses. In this work, we focus on an additional production channel where
the top quark is produced as a component of a monotop state, \ie, in association with missing transverse energy.
While such a system cannot be produced at current collider experiments in the context of
the Standard Model due to loop-suppression and the Glashow-Iliopoulos-Maiani mechanism,
large classes of new physics theories predict its possible observation. Among those,
a first type of scenarios features a monotop signature
arising from the decay of a heavy resonance.
As examples, one finds supersymmetric models with $R$-parity violation in which
a singly-produced top squark decays into a top quark and a long-lived
neutralino giving rise to missing transverse energy~\cite{Berger:1999zt,Berger:2000zk,Desai:2010sq},
models with an extended gauge symmetry featuring leptoquarks than can decay into a particle
pair constituted of a top quark and a neutrino~\cite{Kumar:2013jgb} or hylogenesis scenarios
for dark matter where the top quark is produced together with several
dark matter candidates carrying missing momentum~\cite{Davoudiasl:2011fj}. A second category of beyond the
Standard Model theories leading to monotop production at the LHC with a possibly large rate
exhibits flavor-changing neutral interactions of quarks, and in particular of the top quark, with
fields belonging to a hidden sector. If produced in a collider experiment, these fields are hence manifest through
a significant presence of missing transverse energy. In this theoretical framework, the specific example
of a $Z'$ field dominantly decaying invisibly and coupling to quarks in a flavor-violating fashion
has recently received a particular attention as motivated
by possible solutions to the dark matter problem of the Standard Model~\cite{Kamenik:2011nb,Alvarez:2013jqa}.

While all these models are inspired by the so-called top-down approach for
new physics investigations, an alternative choice focuses on the systematization of
the analysis of a given beyond the Standard Model
effect or signature. It relies on a bottom-up avenue where the Standard Model is extended minimally
through the construction of an effective theory sufficient to reproduce the effect under
consideration. More general monotop investigations have followed this path, concentrating either on
the resonant production mode~\cite{Wang:2011uxa}, on the derived effective four-fermion
interactions that arise when the resonances are heavy~\cite{Morrissey:2005uza, Dong:2011rh},
on monotop production through flavor-changing neutral
interactions~\cite{delAguila:1999ac} or on all cases
simultaneously described via the most possibly general and renormalizable Lagrangian
description~\cite{Andrea:2011ws}. In particular, the parameters of a
part of the non-resonant production mechanisms included in this last Lagrangian, that
is also motivated by a possible explanation for several new physics signals reported
by dark matter direct
detection experiments~\cite{Bernabei:2008yi,Aalseth:2010vx,Aalseth:2011wp,Angloher:2011uu},
have already been constrained by a recent analysis of all Tevatron data by the CDF
collaboration~\cite{Aaltonen:2012ek}. It has been found that light
vector states leading to missing transverse energy in the detector and
interacting in a flavor-violating fashion with the quarks
are restricted to be heavier than about 100~GeV at the 95\% confidence level.

In this work, we revisit and update a pioneering analysis of the LHC sensitivity
to monotops based on the most general and renormalizable dedicated effective field theory.
In this context, the Standard Model is supplemented by
a series of dimension-four operators and new states, the corresponding
Lagrangian encompassing
all possible monotop production modes~\cite{Andrea:2011ws}. However, this analysis
has been first restricted to an estimate of the LHC reach to monotops decaying only
into a purely hadronic state. Additionally, it has been based on parton-level Monte Carlo simulations
of collisions that have been produced during the 2011 run of the LHC collider,
at a center-of-mass energy of 7~TeV. A detailed investigation of the monotop phenomenology
at the LHC, employing the above-mentioned framework,
considering both the hadronic and leptonic decay modes and focusing on the 2012 LHC run,
is therefore still missing.

This paper has the aim to fill this gap and provides,
by means of state-of-the-art Monte Carlo event generation, a first
estimate of the LHC sensitivity to monotops by making use of the 2012 dataset. We base our analysis
on a simulation of 20~fb$^{-1}$ of proton-proton collisions
at a center-of-mass energy of 8~TeV and achieve an accurate background
description by using multiparton matrix-element merging techniques,
precise total cross section predictions for the Standard Model background subprocesses
as well as an advanced modeling of the effects yielded by a CMS-like detector. Within this
setup, we design several search strategies that we think worthwhile to be tested
in the context of the real data by both the ATLAS and CMS experiments. All of them
rely on the amount of missing transverse energy accompanying the top quark which is
expected to be important. In our simplified model approach, this missing transverse
energy is assumed to be due to a neutral
weakly interacting new physics state either of spin zero, one-half or one, and
that is also possibly a candidate for explaining the presence of
dark matter in the Universe. This new particle is then
either long-lived or stable so that it is prevented to leave any visible track in the detector.
In addition, the study of the hadronic monotop signature employs the strength of
the possible top quark reconstruction to reject most of the
background~\cite{Andrea:2011ws,Kamenik:2011nb,Wang:2011uxa}, whereas the extraction
of a leptonic monotop signal is based in on the calculation of the transverse mass
of the two-body system comprised of the lepton issued from the top quark decay and the missing
transverse momentum~\cite{Alvarez:2013jqa}.

This work is organized as follows. In Section~\ref{sec:theory}, we review the
effective field theory that has been constructed to describe all possible monotop production
mechanisms and that we have employed in our analysis. In addition, we
carefully investigate the associated total cross sections as a function
of the different model parameters in order to get a first idea about the mass range
expected to be reachable at the LHC running at a center-of-mass energy of 8~TeV.
We detail in Section~\ref{sec:pheno} the
phenomenological analyses that we have designed in the aim of unveiling new physics
arising through a monotop signature at the LHC. We first describe our Monte Carlo
simulation setup in Section~\ref{sec:simu} and then proceed with the analysis of the
hadronic and leptonic monotop final states in
Section~\ref{sec:hadrmonotops} and Section~\ref{sec:leptmonotops},
respectively. Our conclusions are given in Section~\ref{sec:conclusions}.

\section{Theoretical framework}
\label{sec:theory}

\begin{figure}[t]\centering
  \includegraphics[width=.25\columnwidth]{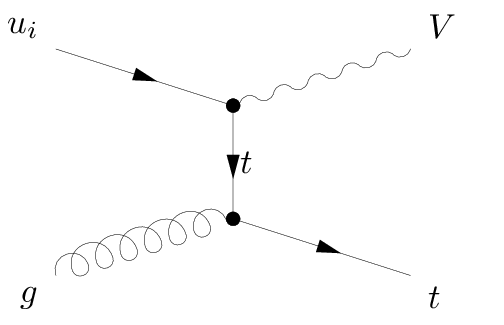}
  \includegraphics[width=.25\columnwidth]{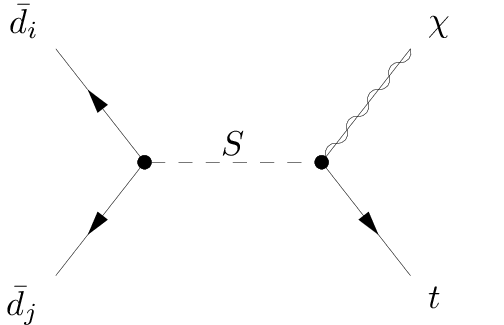}
  \caption{\label{fig:diag} Representative Feynman diagrams leading to a monotop
    signature, either through a flavor-changing neutral interaction of an up-type quark
    with a vector state $V$ (left) or via the
    resonant exchange of a colored
    scalar field $S$ (right). The final state particles $V$ and $\chi$ are
    invisible.}
\end{figure}

As presented in earlier studies, there are two different types
of tree-level processes yielding the production of a monotop
state~\cite{Andrea:2011ws}. This classification depends
on the nature of the particle giving rise to missing transverse energy.
More into details, either the top quark is produced
through a flavor-changing neutral interaction with a lighter quark and a new
invisible bosonic state\footnote{We denote by invisible
any state whose signature in a detector consists of missing transverse
energy.},
as illustrated by the representative Feynman diagram
of the left panel of Figure~\ref{fig:diag}, or in association with
an invisible fermionic state as shown on the representative Feynman
diagram of the right panel of Figure~\ref{fig:diag}.
Starting from the Standard Model, we describe all these monotop
production mechanisms by constructing a simplified model whose Lagrangian
is given, when omitting exotic monotop production mechanisms
involving fields with a spin higher than one or
higher-dimensional operators, by
\be\label{eq:lag}\bsp
 \lag &=\ \lag_{\rm SM} + \lag_{\rm kin} 
  + \bigg[\phi \bar u \Big[a^0_{FC}\!+\!b^0_{FC} \gamma_5 \Big] u \!+\!
     V_\mu \bar u \gmu \Big[a^1_{FC} \!+\! b^1_{FC} \gamma_5 \Big] u  \\
  &+ \varphi \bar d^c
       \Big[a^q_{SR} + b^q_{SR} \gamma_5 \Big] d +
     \varphi \bar u \Big[a^{1/2}_{SR} + b^{1/2}_{SR} \gamma_5 \Big] \chi
  \\ &
   + X_\mu \bar d^c\gmu
       \Big[a^q_{VR} + b^q_{VR} \gamma_5 \Big] d
   + X_\mu \bar u \gmu
       \Big[a^{1/2}_{VR} + b^{1/2}_{VR} \gamma_5 \Big] \chi + 
       {\rm h.c.} \bigg] \ .
\esp\ee
In this approach, the Standard Model Lagrangian $\lag_{\rm SM}$ is first
supplemented by kinetic and gauge interaction terms
for all new states included in the Lagrangian
$\lag_{\rm kin}$. Next, we depict the flavor-changing
associated production of a top quark and a scalar $\phi$ or vector $V$
invisible state by the other terms of the first line of the
equation above, all color and generation indices being understood for clarity.
The strength of the interactions among these two states and a pair
of up-type quarks is modeled via two $3\times 3$
matrices in flavor space $a^{\{0,1\}}_{FC}$
and $b^{\{0,1\}}_{FC}$.
The last two lines of the Lagrangian of Eq.~\eqref{eq:lag} are related
to the second considered class of processes leading to the production of a monotop
system comprised in this case of a top quark
and an invisible fermionic state $\chi$. In this work, we restrict
ourselves to a production mechanism involving the decay of a new colored,
scalar $\varphi$ or vector $X$, resonance
lying in the fundamental representation of the QCD group $SU(3)_c$.
In our notations, the couplings of these new colored fields to down-type quarks are
embedded into the $3\times 3$ matrices
$a^q_{\{S,V\}R}$ and $b^q_{\{S,V\}R}$ while those to the invisible fermion $\chi$ and one
single up-type quarks are given by the three-component vectors
$a^{1/2}_{\{S,V\}R}$ and $b^{1/2}_{\{S,V\}R}$
in flavor space.

Starting from the large series of monotop benchmark models included in the Lagrangian
of Eq.~\eqref{eq:lag}, we consider
a set of simplified scenarios in which all axial couplings vanish,
\be
  b = 0\ .
\ee
Furthermore, we adopt a LHC point of view and only retain interactions possibly enhanced
by parton densities in the proton. We hence fix
\be\label{eq:a}\bsp
  & (a^0_{FC})_{13} = (a^0_{FC})_{31} =
    (a^1_{FC})_{13} = (a^1_{FC})_{31} = a \ , \\
  &  (a^q_{SR})_{12} = \ -(a^q_{SR})_{21} = (a^{1/2}_{SR})_3 =
    (a^q_{VR})_{11} = (a^{1/2}_{VR})_3 = a \ ,
\esp \ee
the other elements of the coupling matrices being set to zero.
Within the above settings, we define four series of scenarios. The first two
are denoted by {\bf SI.s} and {\bf SI.v} and address
monotop production via baryon-number conserving and flavor-changing neutral
interactions in the case where a single scalar ({\bf SI.s}) or vector ({\bf SI.v})
field is added to the Standard Model. These new fields are assumed
to be either stable or to decay invisibly, so that they yield missing transverse energy when produced
in a collider experiment.
The last two sets of scenarios, that we denote by {\bf SII.s} and {\bf SII.v},
focus on the production of a monotop state via the decay of a new scalar ({\bf SII.s})
or vector ({\bf SII.v}) colored
resonance. We further assume, in our simplified setup, that the new
resonances always decay with a branching
ratio equal to unity into a monotop state. While this assumption is in general too
optimistic when one accounts for the dijet decay mode~\cite{Wang:2011uxa}, it
however has the advantage to simplify
the benchmark scenario design as it allows monotop production
to be insensitive to the $a^q_{SR}$ and $a^q_{VR}$ parameters.

\begin{figure}[t!]
  \centering
  \includegraphics[width=.49\columnwidth]{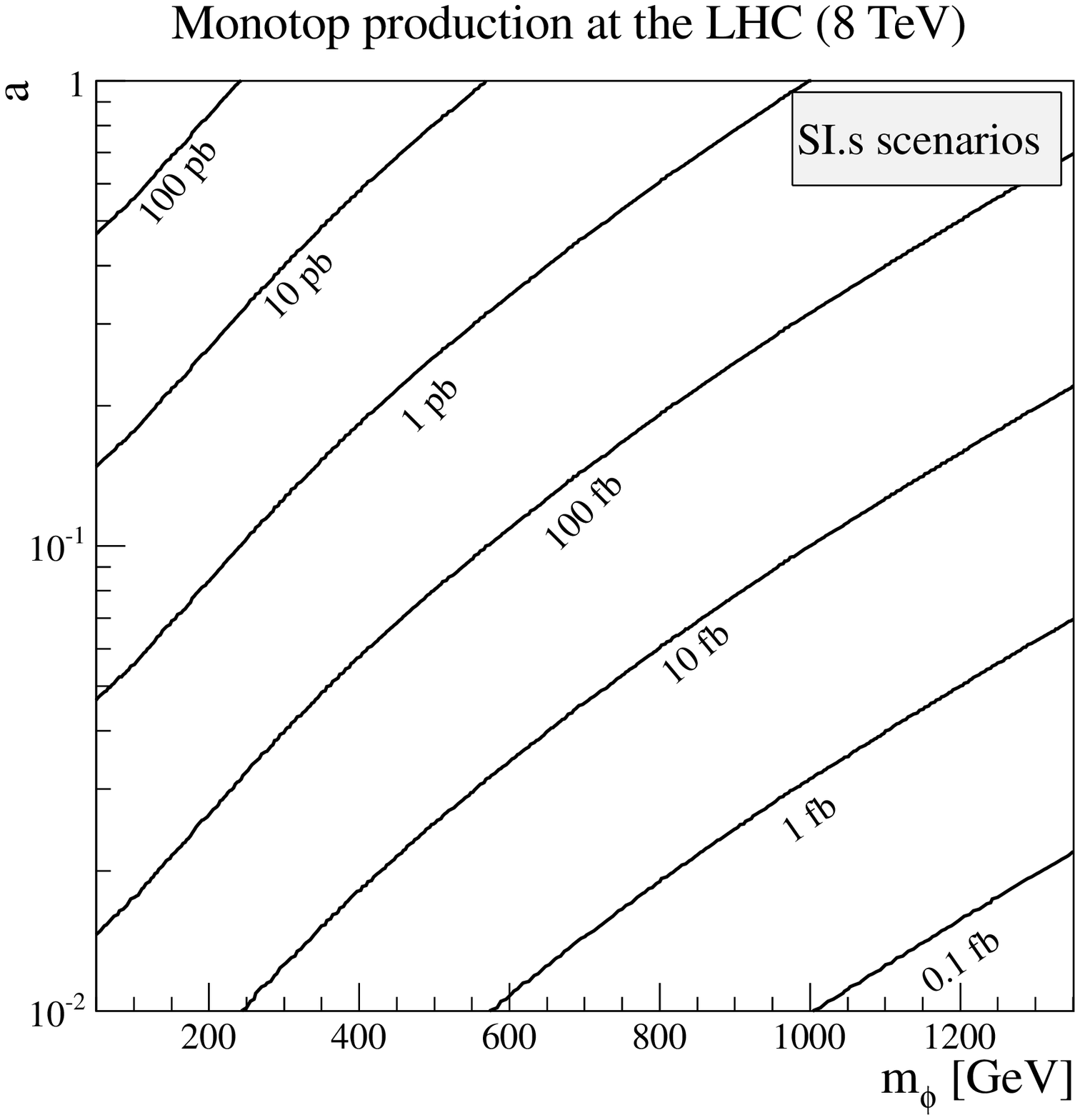}
  \includegraphics[width=.49\columnwidth]{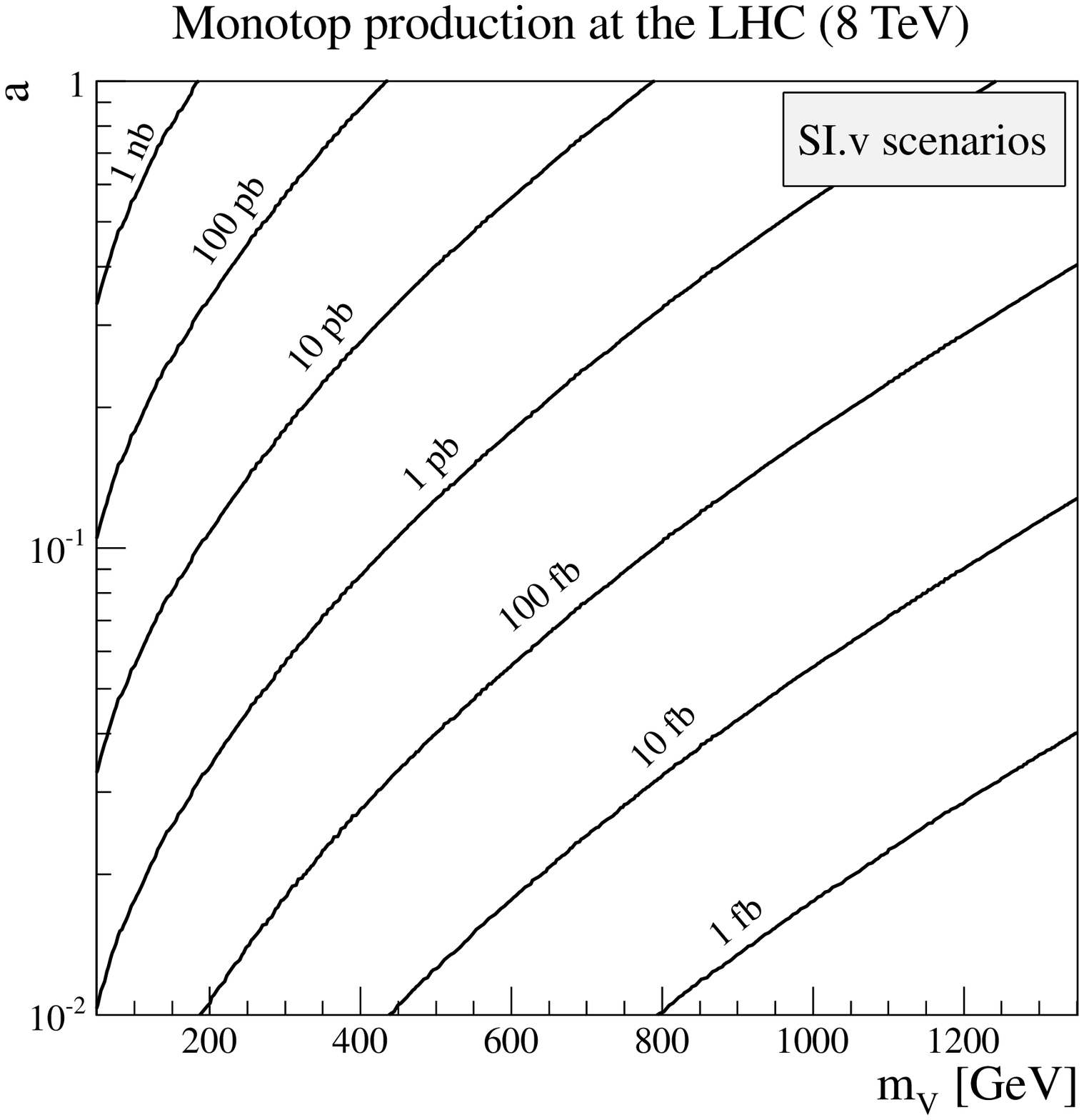}
  \caption{Total cross sections for monotop production at the LHC,
     running at a center-of-mass energy of 8~TeV, for scenarios of type
     {\bf SI.s} (left panel) and {\bf SI.v} (right panel). The results are given as
     a function of the new physics coupling $a$ and the invisible particle mass
     $m_\phi$ (scalar case) and $m_V$ (vector case).}
  \label{fig:xsecS1}
\end{figure}

For each of these classes of scenarios, we estimate monotop production cross sections
at the LHC, running at a center-of-mass energy of $\sqrt{S_h}=8$~TeV, by means
of the QCD factorization theorem. We convolute
the partonic results $\hat \sigma$ as computed by \madgraph~5~\cite{Alwall:2011uj}
with the leading order set of the CTEQ6 fit \cite{Pumplin:2002vw} of
the universal parton
densities $f_{a,b}$ of partons $a,b$ in the proton. Introducing
the longitudinal momentum fractions of the two partons
$x_{a,b} = \sqrt{\tau}e^{\pm y}$, the total cross sections $\sigma$ are
hence given by
\be
 \sigma =
 \sum_{a,b}\int_{\tilde M^2/S_h}^1\!\d\tau
 \int_{-1/2\ln\tau}^{1/2\ln\tau}\d y\
 f_a(x_a,\mu_F^2) \ f_b(x_b,\mu_F^2) \ \hat\sigma(x_a x_b S_h, \mu_R)  \ .
\ee
In our calculations, we fix both the renormalization and factorization scales
$\mu_F$ and $\mu_R$ to the transverse mass of the monotop system
whose production threshold is denoted by $\tilde M$ and we sum the results
related to the production
of a top and an antitop quark in association with missing transverse energy.

In scenarios of class {\bf SI.s} and {\bf SI.v}, a
top quark is produced in association with a bosonic
particle through a flavor-changing interaction,
this bosonic particle being invisible and
giving rise to missing transverse energy.
These two classes of scenarios are described by
two parameters, the mass of the
missing transverse energy particle $m_\phi$ (for {\bf SI.s} scenarios)
and $m_V$ (for {\bf SI.v} scenarios)
and the strength $a$ of the flavor-changing interactions of these particles with
a pair of quarks constituted of one top quark and one up quark, as introduced in
Eq.~\eqref{eq:a}.
In Figure~\ref{fig:xsecS1}, we present monotop production
cross sections as a function of these two parameters for a
scalar (left panel of the figure) and
vector (right panel of the figure) invisible state. We find that
cross sections reaching at least 1~pb can be easily obtained even for
moderate coupling strengths of about $a\approx0.1$. Conversely, such a coupling value
implies that in the context of scenarios of type {\bf SI.s} ({\bf SI.v}),
monotop systems containing an
invisible particle of mass ranging up to about 300~GeV
(500~GeV) could have been abundantly produced during the 2012 LHC run.
Furthermore, scenarios featuring an invisible vector field lead to
a larger
total cross section at the LHC than in the scalar case, when considering
a specific mass and coupling strength. This directly originates
from the vectorial nature of the invisible field included in {\bf SI.v} scenarios.

\begin{figure}[t!]
  \centering
  \includegraphics[width=.49\columnwidth]{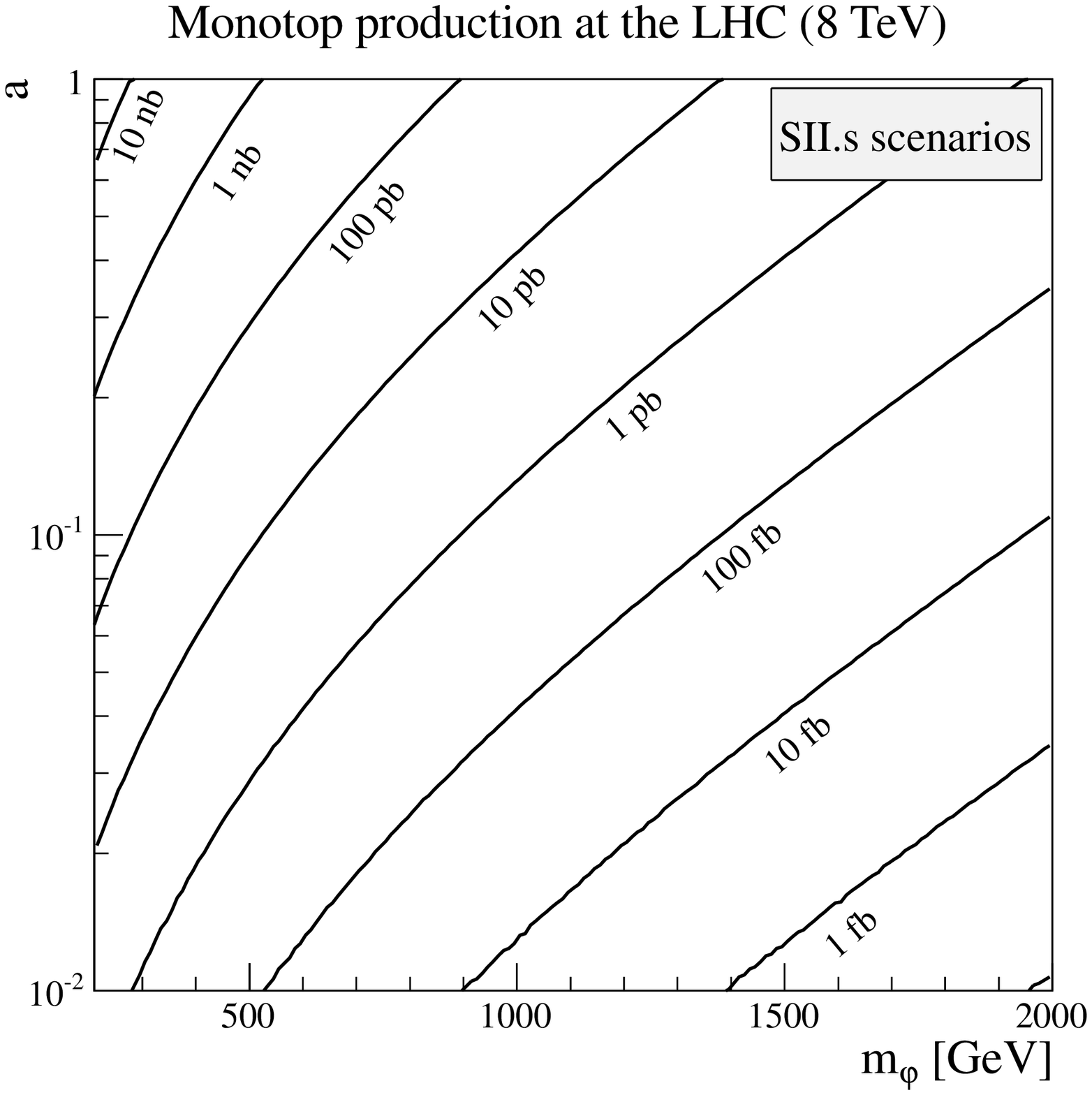}
  \includegraphics[width=.49\columnwidth]{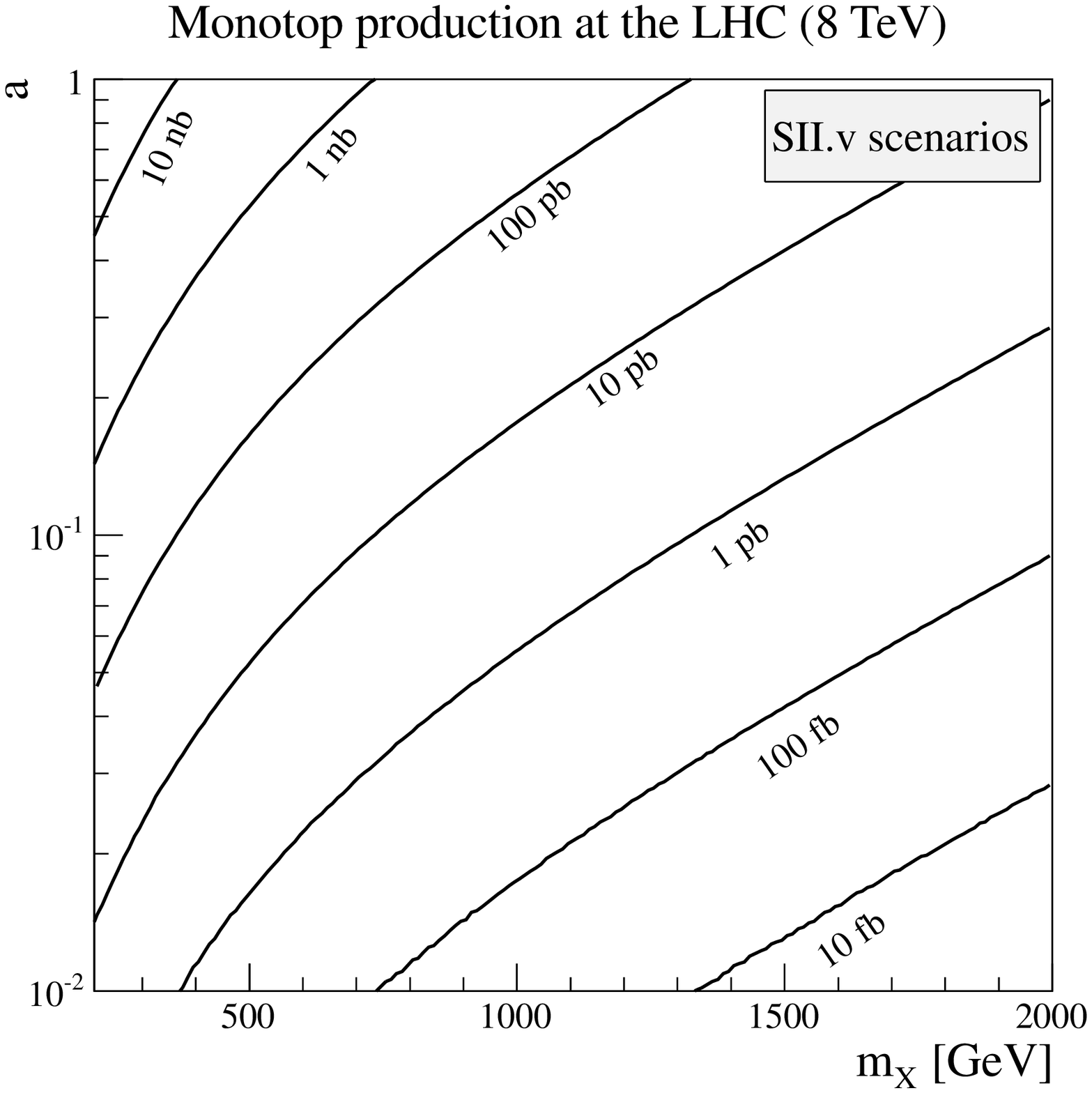}
  \caption{Total cross sections for monotop production at the LHC,
     running at a center-of-mass energy of 8~TeV, for scenarios of type
     {\bf SII.s} (left panel) and {\bf SII.v} (right panel).
     The branching fraction of the resonant state into a monotop system
     is assumed to be unity and the results
     are given as
     a function of the new physics coupling $a$ and of the resonance mass
     $m_\varphi$ (scalar resonance) or $m_X$ (vector resonance).}
  \label{fig:xsecS2}
\end{figure}
In scenarios of type {\bf SII.s} and {\bf SII.v}, we consider that the
monotop state arises from the decay, with a branching fraction equal to
unity, of a new colored scalar $\varphi$ or vector $X$
resonance. In this case, the missing transverse energy is induced by an undetected
fermionic field that we denote by $\chi$. These two classes of scenarios
are described by three parameters, the coupling strength
of the down-type quarks to the new resonance $a$, the resonance mass $m_\varphi$ (for
{\bf SII.s} scenarios) and $m_X$ (for {\bf SII.v} scenarios)
and the mass of the invisible particle $m_\chi$.
Monotop production cross sections are however
independent of the invisible particle mass as they
are equal to the production cross section of the $\varphi$ or $X$ field
multiplied by the subsequent branching ratio into a $t\chi$ pair (being
here set to one). The mass
difference between the resonance and the missing transverse energy particle however alters
the selection efficiency of any monotop search strategy as it is related
to the phase space available
for the resonance decay.
The dependence of the monotop production cross section on the resonance mass and
on the new physics coupling strength $a$ is presented on Figure
\ref{fig:xsecS2} for scenarios involving either
a scalar resonance (left panel of the figure) or
a vector resonance (right panel of the figure).
Cross sections reaching the level of a few pb, allowing for
a large production of monotop events at the LHC, are predicted for
a moderate coupling
strength of $a\approx0.1$ and for resonance masses ranging up to about
1~TeV and 1.3~TeV in the scalar and vector cases, respectively. This
difference finds again its source in the larger cross sections associated with
vector fields.

Monotop production at hadron colliders
can be characterized according to the decay mode of the top quark,
\be\bsp
  &p p \to t + \slashed{E}_T \to b W + \slashed{E}_T \to b j j + \slashed{E}_T \ , \\
  &p p \to t + \slashed{E}_T \to b W + \slashed{E}_T \to b \ell + \slashed{E}_T \ ,
\esp \ee
where $j$ and $b$ denote light and $b$-jets, respectively, $\ell$
a charged lepton and $\slashed{E}_T$ missing transverse energy.
In the next section, we design and investigate two search strategies
associated with each of these two signatures that we dub hadronic and leptonic
monotops. In both cases, we rely on the presence of a large amount of missing
transverse energy carried by the invisible new state. Additionally, we base our analysis,
in the hadronic case, on top quark
reconstruction to reject most of the Standard Model
background~\cite{Andrea:2011ws,Fuks:2012im}, while in the leptonic case,
we employ the $W$-boson transverse mass
to maximize the signal selection efficiency
and significantly reduce the
background~\cite{Berger:1999zt,Berger:2000zk,Alvarez:2013jqa}.

\section{Investigating monotop signatures at the LHC}
\label{sec:pheno}

\subsection{Simulation setup}\label{sec:simu}
In order to investigate monotop production at the LHC, we rely on Monte Carlo
simulations of the 20~fb$^{-1}$ of
collisions that have been produced during the 2012 run at a center-of-mass
energy of 8~TeV.
Event generation for the monotop signal relies on
the implementation of the Lagrangian of
Eq.~\eqref{eq:lag} in the \feynrules\ package~\cite{Christensen:2008py,Alloul:2013bka}
and follows the comprehensive approach for new physics
proposed in Refs.~\cite{Christensen:2009jx} that links a theory
to the study of its phenomenology in a straightforward fashion.
We hence export the simplified theory of Section~\ref{sec:theory}
to the UFO format~\cite{Degrande:2011ua}
and make use of the \madgraph~5 event generator~\cite{Alwall:2011uj}
to simulate parton-level events including the decays of all massive
Standard Model particle. Using the QCD factorization
theorem, tree-level matrix elements are convoluted in this way
with the leading order set of the CTEQ6 parton
density fit~\cite{Pumplin:2002vw}, the renormalization and
factorization scales being set to the transverse mass of the produced heavy
particles. Concerning
the background, we directly employ the built-in Standard Model implementation of
\madgraph~5.

\begin{table}[!t]
\begin{center}
\begin{tabular}{l || c c}
Process & $~~\sigma$ [pb]$~~$& $~~N~~$\\
\hline
\hline
  $W(\to \ell\nu) + \text{jets}$ & 35678    & $2.56\cdot 10^8$\\
  $\gamma^{*}/Z(\to 2 \ell/2 \nu) + \text{jets}$ & 10319 & $4 \cdot 10^7$\\
\hline 
  $t\bar{t} (\to 6 \text{jets}) \!+\! \text{jets}$& 116.2& $8\cdot 10^6$ \\
  $t\bar{t} (\to 4 \text{jets} \ 1\ell\ 1\nu) \!+\! \text{jets}$&  112.4& $9\cdot 10^6$ \\
  $t\bar{t} (\to 2\text{jets} \ 2 \ell\ 2\nu) \!+\! \text{jets}$&  27.2& $3\cdot 10^6$
    \\
\hline 
  Single top + jets [$t$-channel, incl.] & 87.2 & $6\cdot10^6$ \\
  Single top + jets [$tW$-channel, incl.]& 22.2 & $1\cdot10^6$ \\
  Single top + jets [$s$-channel, incl.] & 5.55 & $8\cdot10^5$ \\
\hline
  $t\bar{t}W  + \text{jets}$ [incl.]  & 0.25 & $3\cdot10^4$\\
  $t\bar{t}Z  + \text{jets}$ [incl.]  & 0.21 & $5\cdot10^4$\\
  $t/\bar{t} + Z + j + \text{jets}$ [incl.] & 0.046& $3\cdot10^5$ \\
  $t\bar{t}WW + \text{jets}$ [incl.] & 0.013  & $2\cdot10^3$\\
  $t\bar{t}t\bar{t} + \text{jets}$ [incl.] & $7 \cdot 10^{-4}$& $10^{3}$ \\
\end{tabular}
\hspace{1cm}
\begin{tabular}{l || c c}
Process & $~~\sigma$ [pb]$~~$& $~~N~~$\\
\hline
\hline
  $WW(\to 1\ell\ 1\nu\ 2\text{jets}) \!+\! \text{jets}$ & 24.3 & $3\cdot10^6$\\
  $WW(\to 2\ell\ 2\nu) + \text{jets}$ & 5.87 & $8\cdot10^5$ \\
\hline
  $WZ(\to 1\ell\ 1\nu\ 2\text{jets}) + \text{jets}$ & 5.03 & $5\cdot10^5$ \\
  $WZ(\to 2\nu\ 2\text{jets}) + \text{jets}$   & 2.98 & $3\cdot10^5$ \\
  $WZ(\to 2\ell\ 2\text{jets}) + \text{jets}$  & 1.58 & $2\cdot10^5$ \\
  $WZ(\to 1\ell\ 3\nu) + \text{jets}$          & 1.44 & $2\cdot10^5$ \\
  $WZ(\to 3\ell\ 1\nu)+ \text{jets}$           & 0.76 & $2\cdot10^6$ \\
\hline
  $ZZ(\to 2\nu\ 2\text{jets})+ \text{jets}$   & 2.21 & $3\cdot10^5$ \\
  $ZZ(\to 2\ell\ 2\text{jets})+ \text{jets}$  & 1.18 & $1.5\cdot10^4$ \\
  $ZZ(\to 4\nu)+ \text{jets}$                 & 0.63 & $1\cdot10^5$ \\
  $ZZ(\to 2\nu\ 2\ell)+ \text{jets}$          & 0.32 & $4\cdot10^4$ \\
  $ZZ(\to 4\ell)+ \text{jets}$                & 0.17 & $4\cdot10^4$ \\
\end{tabular}
\caption{Simulated background processes given together with
the related cross section $\sigma$ and
number of generated events $N$. The
background contributions are split
according to the massive state decays, $\ell$ standing equivalently for
electrons, muons and taus, $\nu$ for any neutrino, and $j$ for a jet.
The notation \textit{incl.} indicates
that the sample is inclusive in the decays of the heavy particles.}
\label{tab:xsec}
\end{center}
\end{table}

Parton-level events are further integrated into a full hadronic
environment by matching hard scattering matrix elements with the parton
showering and hadronization infrastructure provided by the
\pythia~6~\cite{Sjostrand:2006za} and
\pythia~8~\cite{Sjostrand:2007gs} programs for the background
and new physics processes, respectively\footnote{The color structures
included in the Lagrangian of Eq.~\eqref{eq:lag} being not all fully supported
by \pythia~6, we have employed the newer version of the program for signal
event generation.}. In the case of the background,
we employ the $k_T$-MLM prescription~\cite{Mangano:2006rw} to
merge event samples described by matrix elements containing
additional jets. While we allow
events associated with the production
of a single gauge boson to contain
up to four extra jets, this number is restricted to two
for any other process.
In addition, tau lepton decays possibly arising from the decays of heavier states
are handled with the \tauola\ program \cite{Davidson:2010rw} and
detector effects are accounted for by using
the \delphes~2 package~\cite{Ovyn:2009tx} together with
the recent CMS detector description of Ref.~\cite{Agram:2013koa}.
Finally, jet reconstruction is performed with the \fastjet\
library~\cite{Cacciari:2011ma}, using an anti-$k_{T}$
algorithm with a radius parameter set to $R=0.4$~\cite{Cacciari:2008gp}, and
the reconstructed events are analyzed within
the \madanalysis~5~framework~\cite{Conte:2012fm,Conte:2013mea}.

Background events are reweighted so that they
reproduce total cross sections
including higher-order corrections when available, making use in this
case of the CT10 next-to-leading order (NLO) parton density fit~\cite{Lai:2010vv}
for the predictions. The values that we have used
are summarized in Table~\ref{tab:xsec} and presented together with the number
of generated events, after distinguishing the different final states
arising from the decays of the weak gauge bosons and
top quarks. More into details, we have normalized
all events originating from the production of a single gauge boson
to the next-to-next-to-leading order (NNLO) accuracy~\cite{Gavin:2012sy,Gavin:2010az},
top-antitop and single top events to the NLO one but
including genuine NNLO contributions~\cite{Aliev:2010zk,Kidonakis:2012db},
diboson, $ttW$ and $ttZ$ events to the pure NLO~\cite{Campbell:1999ah,%
Campbell:2011bn,Campbell:2012dh}, while we use
the leading-order results as computed by \madgraph~5
for all other simulated Standard Model processes.
Moreover, QCD multijet contributions have been omitted as their correct treatment
requires data-driven methods.
We have instead chosen to resort on available experimental studies to ensure a good control
of this source of background by designing appropriate event selection strategies (see
the next subsections).

\subsection{Hadronic monotops at the LHC}\label{sec:hadrmonotops}

In order to probe hadronic monotops, we build an approach whose
cornerstone
benefits from the possible reconstruction of the hadronically decaying
top quark. Event preselection
involves the presence of exactly one jet tagged as originating from the
fragmentation of a $b$-quark and two or three light jets, this last criterion
allowing us to make use of monotop events featuring initial or final state
radiation. This choice is motivated by an increase of the sensitivity $s$
to monotops, $s$ being defined as the significance $S/\sqrt{S+B}$ where
$S$ and $B$ are respectively the number of signal and background events
after all selections.
The selected jet candidates
are required to lie within the detector geometrical acceptance, \ie,
with a pseudorapidity satisfying $|\eta^j| < 2.5$, and we impose that
their transverse momentum $p_T$ is greater than 30~GeV and 50~GeV
for light and $b$-tagged jets, respectively. Moreover,
we demand that for any considered jet,
the ratio between the hadronic and electromagnetic calorimeter deposits
is larger than 30\%. Additionally, we veto any event
containing at least one identified isolated electron or muon
whose transverse momentum and pseudorapidity satisfy
$p_T^\ell > 10$~GeV and $|\eta^\ell| < 2.5$, respectively.
We define lepton isolation on
the basis of a variable $I_{\rm rel}$ corresponding to
the amount of transverse energy, evaluated
relatively to $p_T^\ell$,
present in a cone of radius $R \!=\! \sqrt{\Delta\varphi^2 +
\Delta\eta^2} \!=\! 0.4$ centered on the lepton, $\varphi$ being the azimuthal angle with
respect to the beam direction. We constrain this quantity
such as $I_{\rm rel} < 20\%$.

After this preselection, we expect about $8\cdot 10^5$ background events,
although this number does not account for
non-simulated multijet contributions.
A good fraction of these events (35\%) are
issued from the production of a leptonically decaying
$W$-boson in association with jets, the charged lepton being either
too soft ($p_T^\ell < 10$~GeV), non-isolated, or
lying outside the detector acceptance ($|\eta^\ell| >
2.5$).
The next-to-leading background component (25\% of the preselected events)
is made up of top-antitop events. For half of them,
both top quarks decay
hadronically whereas for the other half, only one of the top quarks
decays hadronically and the other one gives rise to
a non-reconstructed lepton.
The rest of the background finds its origin in
single top production (20\%), in the associated production of
an invisibly-decaying $Z$-boson with jets (15\%) while
any other contribution,
such as events originating from diboson or rarer Standard Model processes,
is subdominant.

\begin{figure}
  \begin{center}
    \includegraphics[width=0.49\columnwidth]{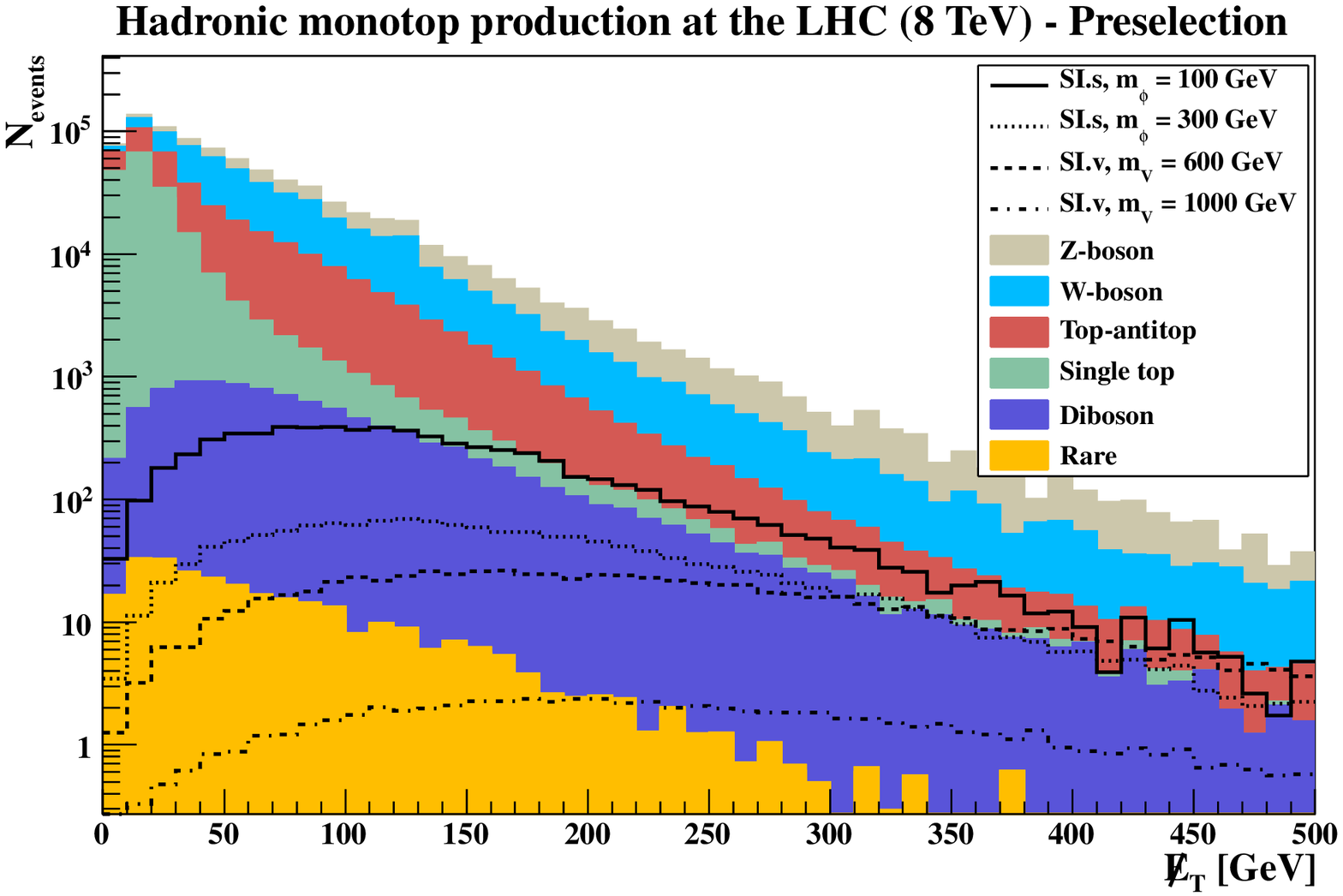}
    \includegraphics[width=0.49\columnwidth]{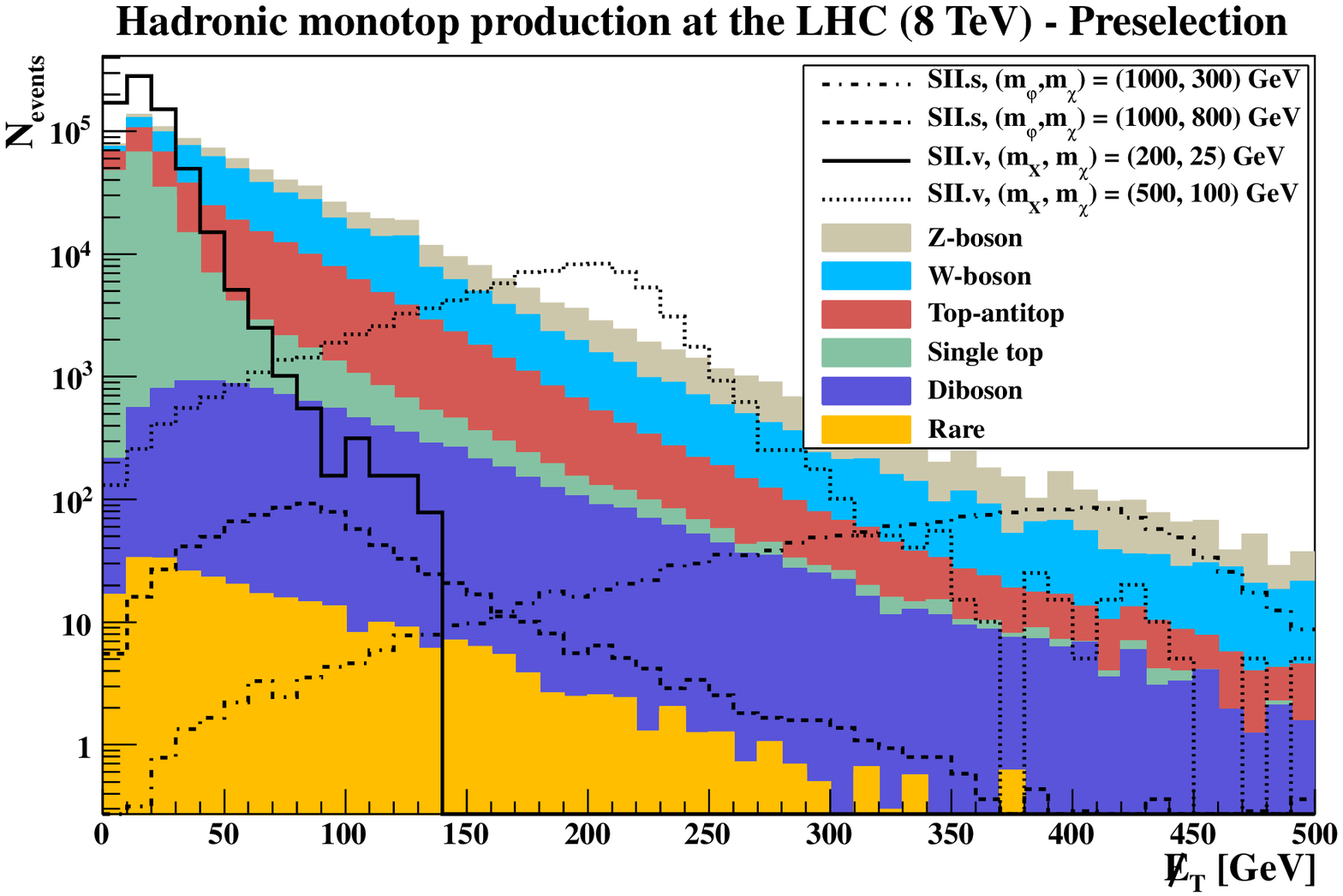}
    \caption{Missing transverse energy distributions after preselecting
    events with one single $b$-tagged jet, two or three light
    jets and no isolated electron or muon. We represent separately (and stacked) the various
    contributions to the Standard Model expectation, to which we superimpose
    predictions for eight representative signal
    scenarios of class {\bf SI} (left panel) and {\bf SII} (right panel),
    the new physics coupling strengths being each time fixed to $a = 0.1$.}
    \label{fig:hmonotops_met}
  \end{center}
\end{figure}

In addition to the presence of a single top quark,
monotop signal events contain a significant amount
of missing transverse energy $\slashed{E}_T$ carried by the invisible new state.
Calculating $\slashed{E}_T$ as the norm of the sum of the transverse momenta
of all the visible objects, we present on Figure~\ref{fig:hmonotops_met}
its distribution for the
different background contributions and a few representative signal scenarios.
On the left panel of the figure,
we focus on scenarios of class
{\bf SI} where the monotop state arises from
a flavor-changing neutral interaction. We consider lighter scalar invisible
particles ({\bf SI.s} scenarios) with
$m_\phi=100$~GeV and $m_\phi=300$~GeV and heavier vector invisible
particles ({\bf SI.v} scenarios) with
$m_V = 600$~GeV and $m_V=1000$~GeV.
Each signal distribution exhibits a peak whose
position depends on the invisible particle mass and a tail
extending, with a gently decreasing slope, to missing transverse energy values larger
than 500~GeV.
For new states heavier than about 100~GeV, few signal events are expected
in the parameter space region where the bulk of the Standard Model events lies,
so that a key requirement on the missing transverse energy would simultaneously allow for a
good background rejection and an important signal
selection efficiency.

On the right panel of Figure~\ref{fig:hmonotops_met},
we consider four representative {\bf SII} scenarios where the monotop
signature arises from the decay of a colored resonance.
Two of these feature a heavy scalar resonant state
of mass $m_\varphi = 1000$~GeV ({\bf SII.s} scenarios) and
a monotop system constituted of a top quark and an invisible
fermion being either quite light ($m_\chi=300$~GeV)
or rather heavy ($m_\chi=800$~GeV). In the first case, the available
phase space for the resonance decay is important while in the second
case, the monotop state has to be produced almost at threshold.
The last two scenarios focus on vector resonances (type {\bf SII.v}). For
one of them, the resonant state is rather light ($m_X=200$~GeV) and the mass of
the invisible fermion is fixed to $m_\chi=25$~GeV, while for the other one,
the resonance is heavier ($m_X = 500$~GeV) and the invisible state is
still moderately light ($m_\chi = 100$~GeV). All missing transverse energy distributions
present a typical resonant behavior with an edge, distorted
due to detector effects, at a $\slashed{E}_T$ value depending both
on the resonance and invisible particle masses. When the monotop system has
to be produced close to threshold, monotop events are not expected to
contain a large quantity of missing transverse energy so that any $\met$
requirement with a good Standard Model background rejection
has consequently
a poor signal efficiency. When the mass difference between the two new states
increases, the edge of the $\slashed{E}_T$ spectrum moves to larger
values so that the sensitivity to such scenarios is expected to be more important.

From those considerations, we
design two different hadronic monotop search strategies. The first one
is dedicated to the low mass region of the parameter space
with a lower selection threshold on the missing transverse energy, whereas
the second one is based on a harder selection
more sensible to the high mass region. We choose
\be
  \text{either}\qquad \slashed{E}_T > 150~\text{GeV}\qquad\text{or}\qquad
  \slashed{E}_T > 250~\text{GeV}\ .
\ee
Such requirements on the missing transverse energy, together with the current
preselection criteria on jets, are also expected to imply
a good control of the non-simulated multijet
background~\cite{Andrea:2011ws,daCosta:2011qk,Collaboration:2011ida}.
After imposing the looser $\slashed{E}_T > 150$~GeV
criterion\footnote{Such a low missing transverse energy requirement
might however be challenging with respect to $\met$-only trigger efficiencies
in the context of a real experiment.},
about 45000 background events
survive and are mostly consisting of events related to the production of an invisibly
decaying $Z$-boson (43\%), a $W$-boson (37\%) or a top-antitop pair
(15\%) with jets. With the tighter missing transverse energy requirement
$\slashed{E}_T > 250$~GeV, most of the Standard Model background is rejected and
only about 8000 events, originating from
the production of a $Z$-boson (53\%),
a $W$-boson (33\%) or a top-antitop pair (8\%) in association with jets, remain.

The next steps of the selection strategy are based on the top quark
reconstruction from the final state jets and on
the kinematical configuration of the signal
events, the missing transverse momentum and the reconstructed top quark being mainly
back-to-back. Two of the selected light jets $j_1$ and $j_2$
are hence imposed to have an invariant
mass $M_{j_1j_2}$ compatible with the mass of a $W$-boson,
\be
  M_{j_1j_2}\ \in\ \ ]50, 105[~\text{GeV} \ .
\ee
In the case of events with three selected light jets,
$M_{j_1 j_2}$ is defined as the invariant mass
of the dijet system whose invariant mass is the closest
to the $W$-boson mass.
In addition, we require the two-momentum of the leading jet
$\vec{p}(j_1)$ to be non-collinear with the missing transverse momentum $\vec{\slashed{p}}_T$
defined as the opposite of the
sum of the transverse momenta of all visible objects,
\be
  \Delta\varphi \Big(\vec{\slashed{p}}_T, \vec{p} (j_1)\Big) \in\ \ ]0.5, 5.75[ \ .
\ee
Taking into account the selected $b$-tagged jet, we fully reconstruct the top quark
and demand that its two-momentum $\vec{p}(t)$
is well separated from the missing two-momentum,
\be
  \Delta\varphi \Big(\vec{\slashed{p}}_T, \vec{p}(t)\Big) \in \ \ ]1, 5[ \ .
\ee

\begin{figure}
  \begin{center}
    \includegraphics[width=0.49\columnwidth]{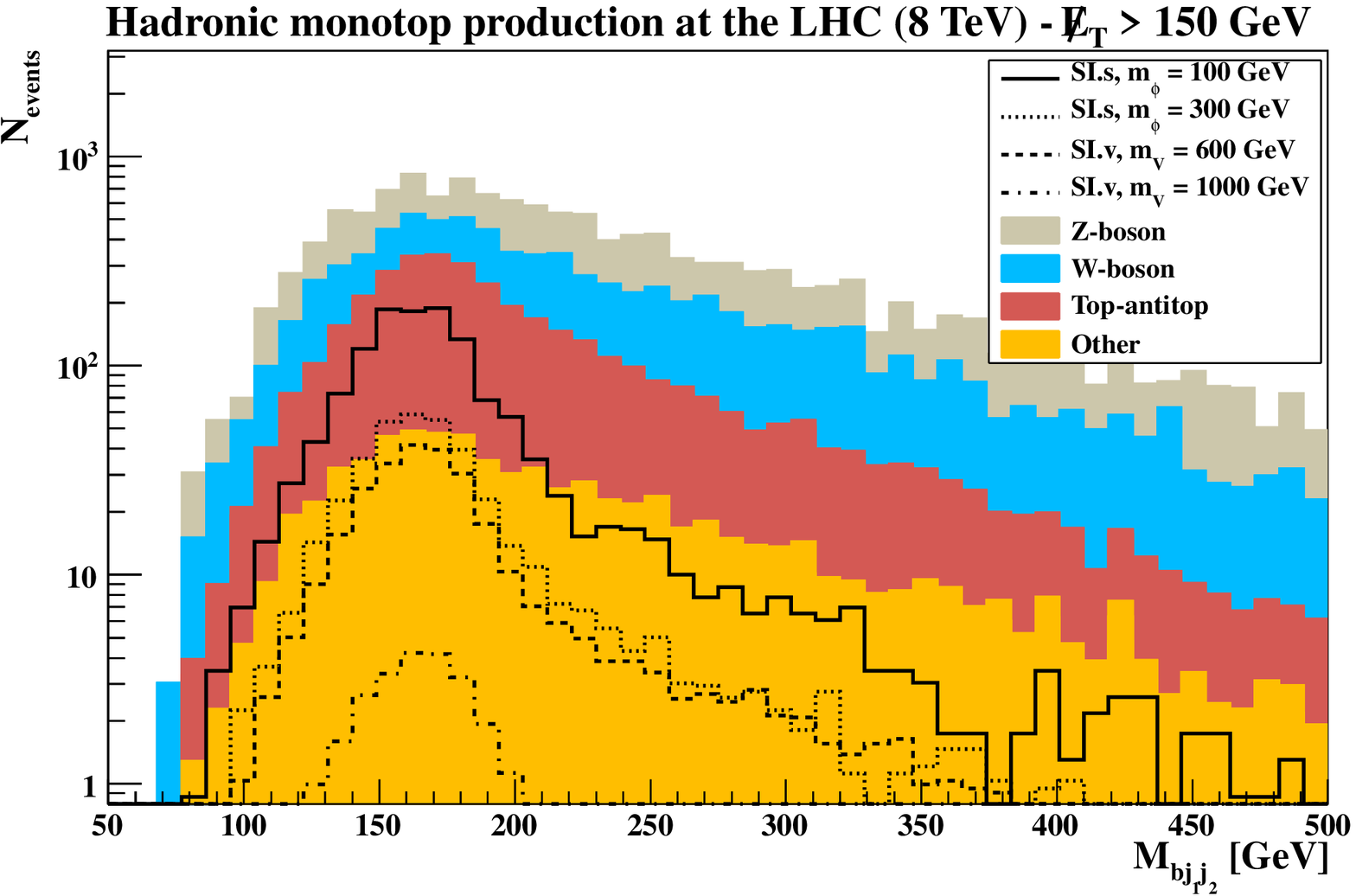}
    \includegraphics[width=0.49\columnwidth]{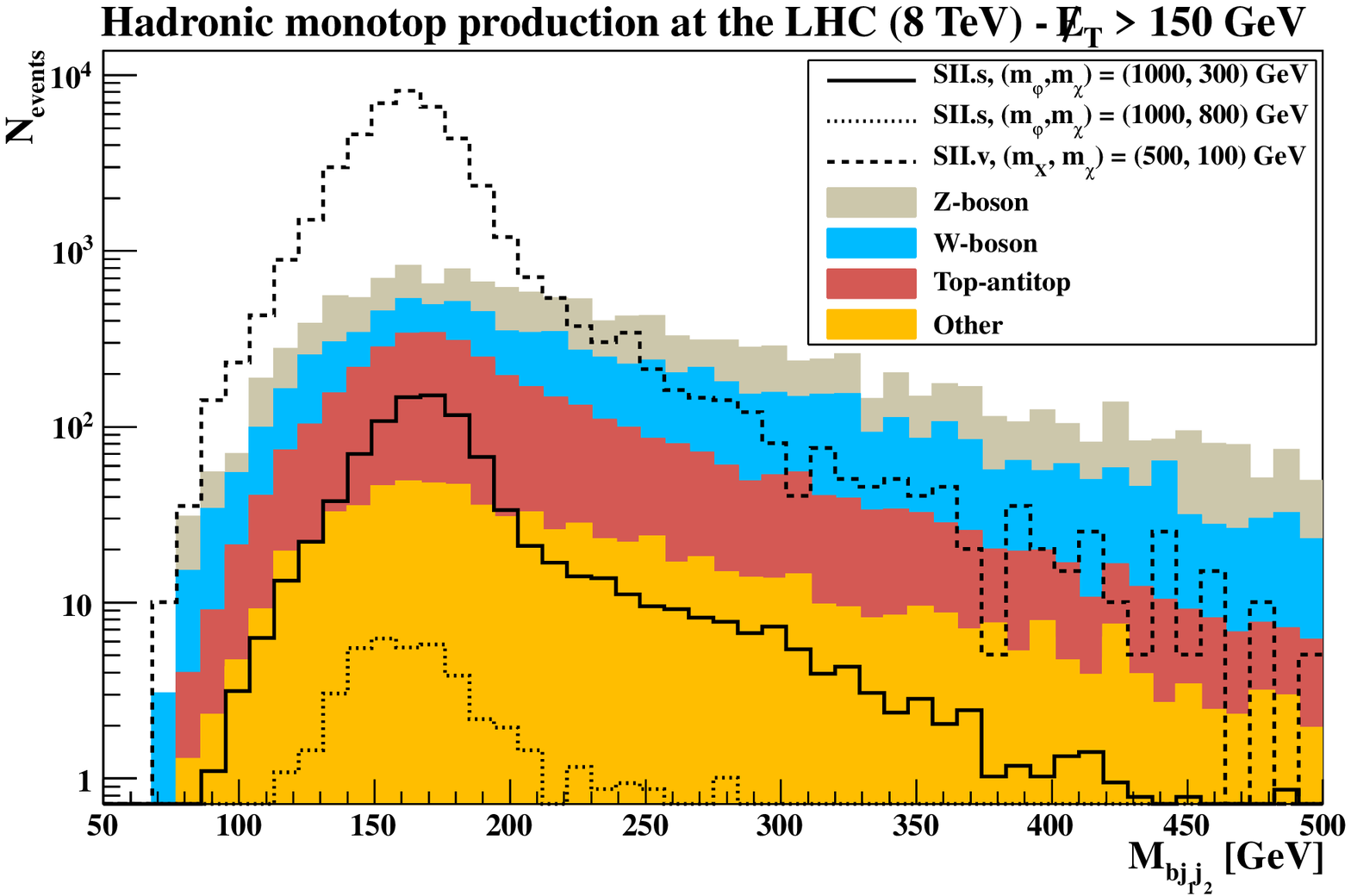}
    \caption{Invariant mass distribution of the reconstructed top quark after all
    requirements described in the text and a loose missing transverse energy
    selection of $\slashed{E}_T > 150$~GeV. We present separately (and stacked) the various
    contributions to the Standard Model expectation to which we superimpose
    predictions for eight representative signal
    scenarios of class {\bf SI} (left panel) and {\bf SII} (right panel),
    the coupling strengths being fixed to $a = 0.1$.}
    \label{fig:monotopmjjb_a}
    \vspace{.2cm}
    \includegraphics[width=0.49\columnwidth]{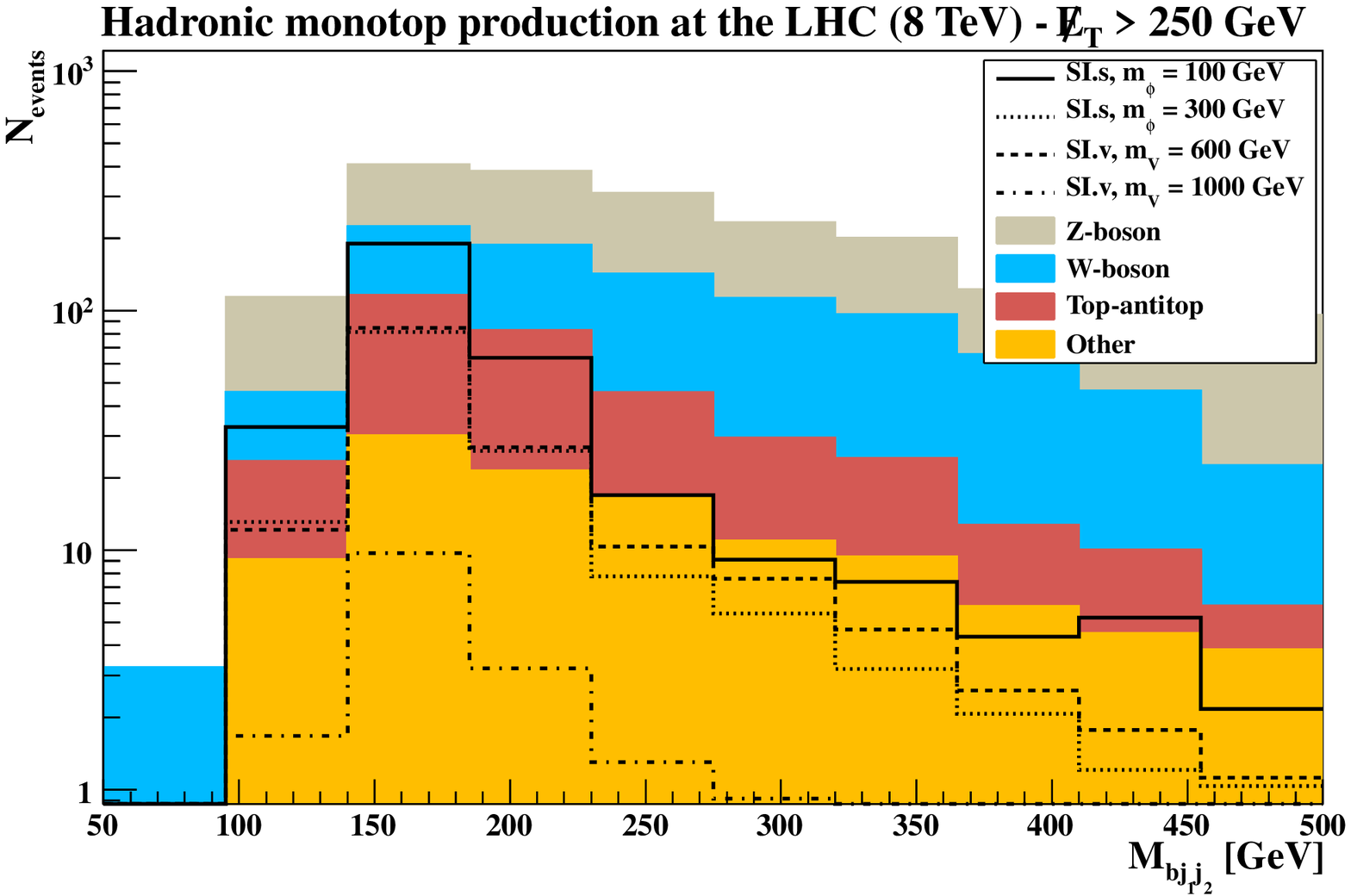}
    \includegraphics[width=0.49\columnwidth]{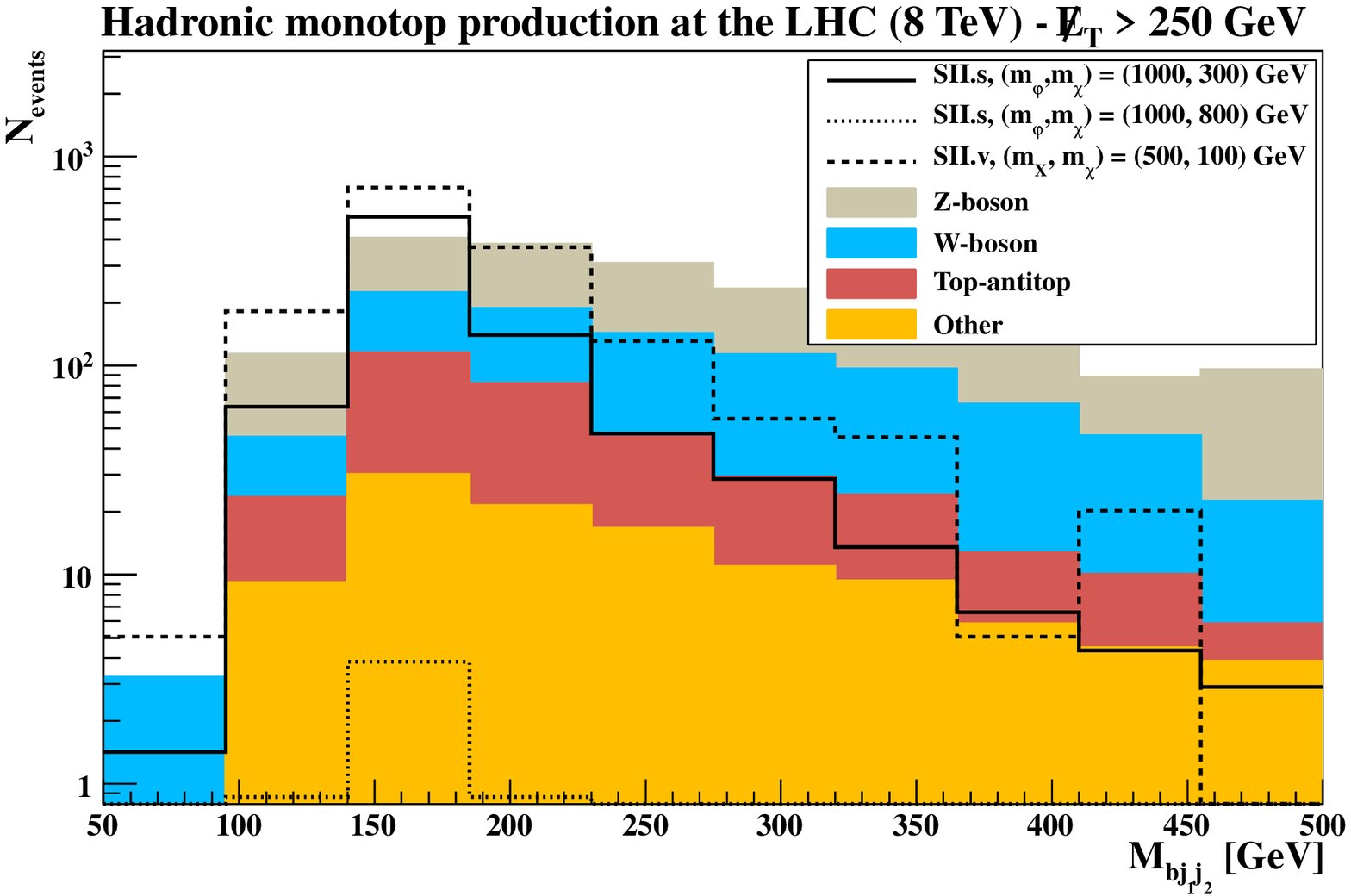}
    \caption{Same as in Figure~\ref{fig:monotopmjjb_a} but for
    a missing transverse energy
    requirement of $\slashed{E}_T~>~250$~GeV.}
    \label{fig:monotopmjjb_b}
  \end{center}
\end{figure}

At this stage of the analysis, the background is comprised of about
15000 (2000) events when one imposes that $\slashed{E}_T > 150$~GeV (250~GeV),
composed of 40\% (52\%), 33\% (31\%) and 22\% (11\%)
of events related to
the production of a $Z$-boson, a $W$-boson and a top-antitop pair,
respectively. The results of the two analysis strategies
are illustrated on Figure~\ref{fig:monotopmjjb_a} (loose
$\met$ requirement) and
Figure~\ref{fig:monotopmjjb_b} (tight $\met$ requirement)
where we present the spectrum in the invariant
mass of the reconstructed top quark $M_{bj_1j_2}$
for the dominant background contributions and a few representative signal scenarios
of class {\bf SI} (left panel of the figures) and {\bf SII} (right panel of the figures).
Since the bulk of signal
events are populating bins with a value around the top mass with a narrower spread than for
the background, whose spectrum also exhibits a continuum
extending to larger $M_{bj_1j_2}$ values,
we enforce
\be
  M_{bj_1j_2}\ \in\ \ ]140, 195[~\text{GeV} \ .
\ee
This allows for a good signal selection efficiency and an important
rejection of the background contamination.

\begin{table}
  \begin{center}
    \begin{tabular}{c||c|c}
         Process & $N_{\rm ev}$, $\slashed{E}_T > 150$~GeV & $N_{\rm ev}$, $\slashed{E}_T > 250$~GeV\\
       \hline
       \hline
           $Z$-boson production & $1411 \pm 38$ & $210 \pm 15$ \\
           $W$-boson production & $1064 \pm 33$ & $148 \pm 12$ \\
           Top-antitop pair production & $1486 \pm 39$ & $105 \pm 10$ \\
           Other background contributions& $262 \pm 15$ & $34.7 \pm 5.9$ \\
       \hline
       Total background & $4223 \pm 65$ & $497 \pm 22$\\
       \hline
     \hline
     {\bf SI.s}, $m_\phi =  100$~GeV & $885  \pm 29$  & $212 \pm 15$ \\
     {\bf SI.s}, $m_\phi =  300$~GeV & $268 \pm 16$  & $92.0 \pm 9.5$\\
     {\bf SI.v}, $m_V =  600$~GeV     & $191  \pm 14 $ &  $95.6 \pm 9.7$ \\
     {\bf SI.v}, $m_V = 1000$~GeV     & $ 19.7 \pm 4.3$ & $11.0 \pm 3.3$ \\
     {\bf SII.s}, $m_\varphi = 1000$~GeV, $m_\chi = 300$~GeV & $664  \pm  25$ & $581 \pm  23$ \\
     {\bf SII.s}, $m_\varphi = 1000$~GeV, $m_\chi = 800$~GeV & $29.4 \pm 5.4$ & $4.1 \pm 2.0$\\
     {\bf SII.v}, $m_X =  200$~GeV, $m_\chi =  25$~GeV & $\approx 0$ & $\approx 0$   \\
     {\bf SII.v}, $m_X =  500$~GeV, $m_\chi = 100$~GeV & $31047 \pm 171$ & $334 \pm 18$ \\
    \end{tabular}
     \caption{\label{tab:hmonotops}
  Number of expected events ($N_{\rm ev}$), after applying all the selections described in the text,
  for 20 fb$^{-1}$ of LHC collisions at a center-of-mass energy of 8~TeV,
  given together with the associated statistical uncertainties. We present separately the different
  contributions to the Standard Model background and results for the eight
  representative signal scenarios introduced in this section. The new physics coupling parameter is
  set to $a=0.1$.}
  \end{center}
\end{table}

\begin{figure}
  \begin{center}
     \includegraphics[width=0.49\columnwidth]{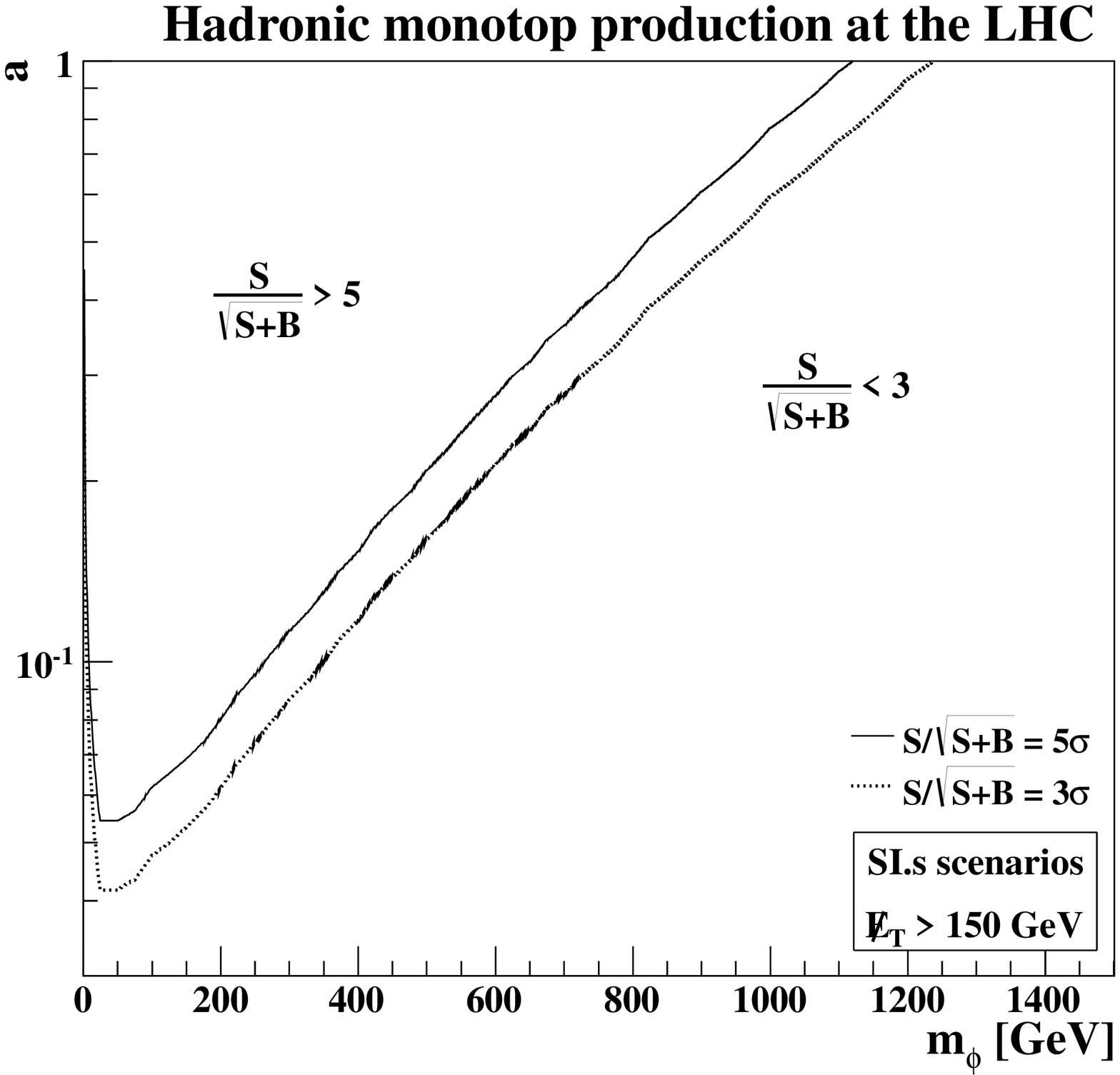}
     \includegraphics[width=0.49\columnwidth]{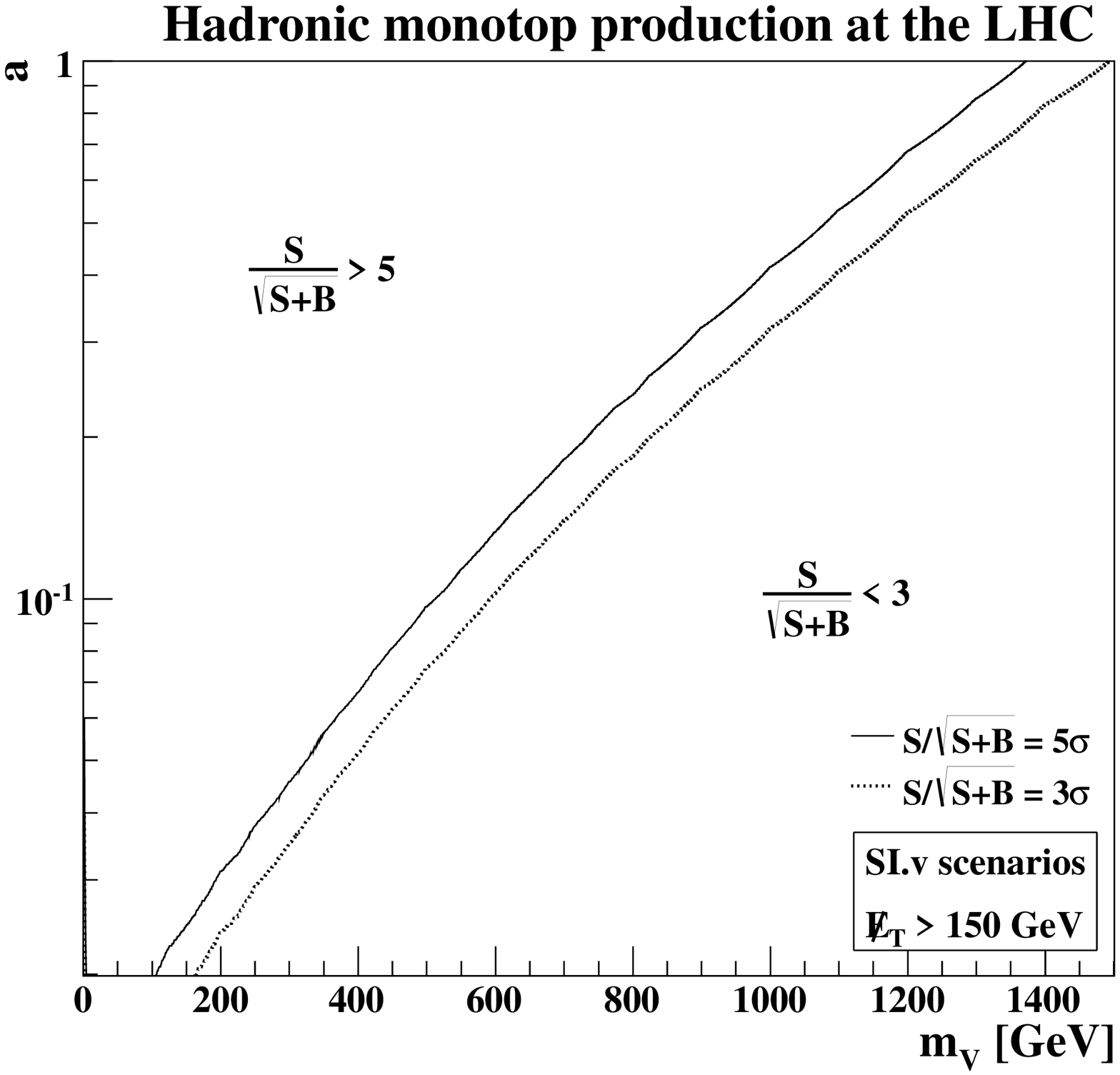}
    \caption{LHC sensitivity to hadronic
    monotop production in the context of scenarios of
    class {\bf SI} where the Standard Model is supplemented by
    a scalar (left panel) or vector (right panel)
    invisible particle, for 20~fb$^{-1}$ of
    proton-proton collisions at a center-of-mass energy of 8~TeV. The sensitivity
    is calculated as the ratio $S/\sqrt{S+B}$ where $S$ and $B$ are the number of signal
    and background events surviving all selections presented in the text.
    The results are given in $(m_\phi,a)$ and $(m_V,a)$ planes for scenarios of class
    {\bf SI.s} and {\bf SI.v}, respectively, and we focus
    on the loose $\slashed{E}_T$ requirement which leads to better sensitivity in the case
    of {\bf SI} scenarios.}
    \label{fig:monotopsign_b}\vspace{.5cm}
     \includegraphics[width=0.49\columnwidth]{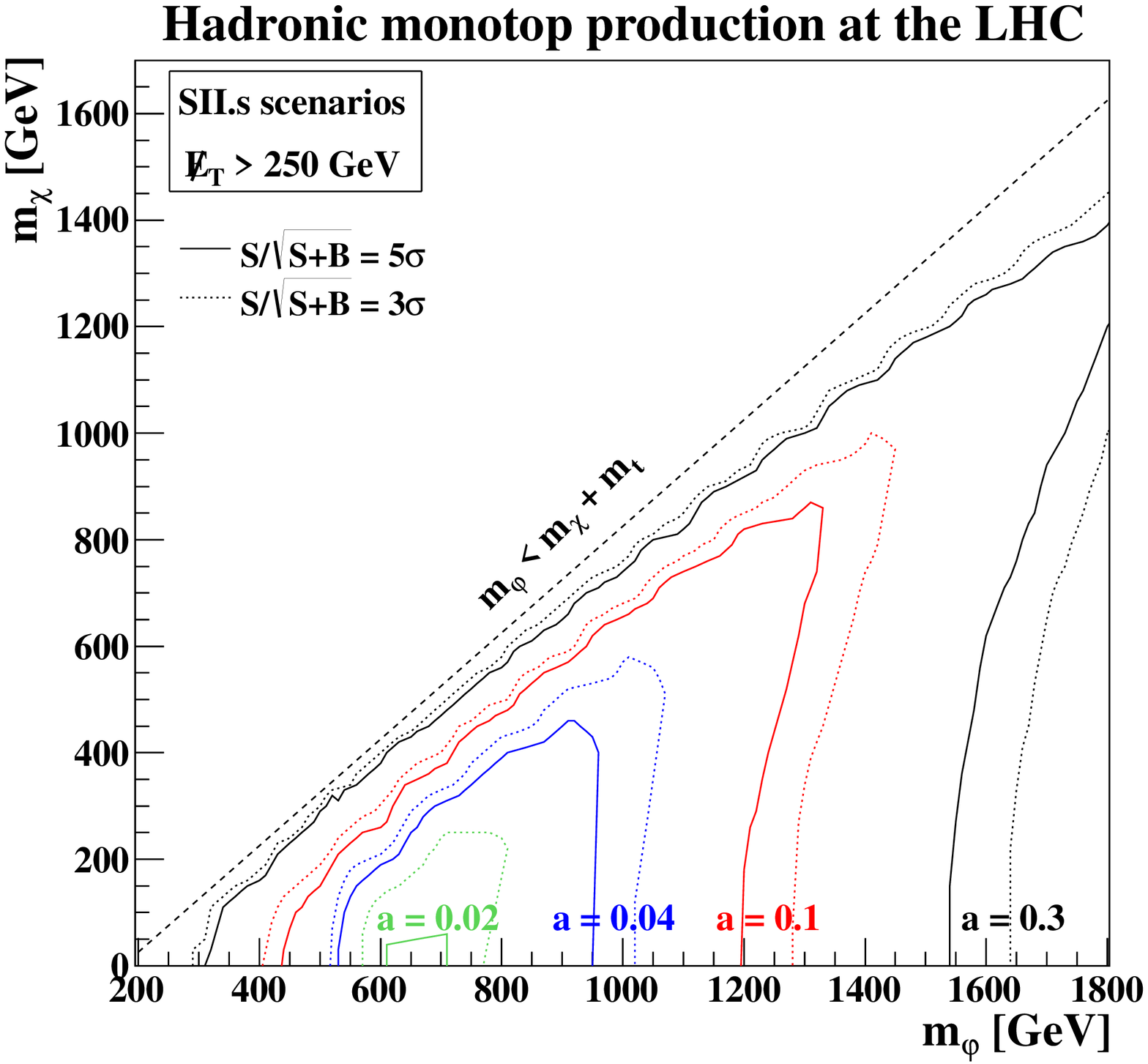}
     \includegraphics[width=0.49\columnwidth]{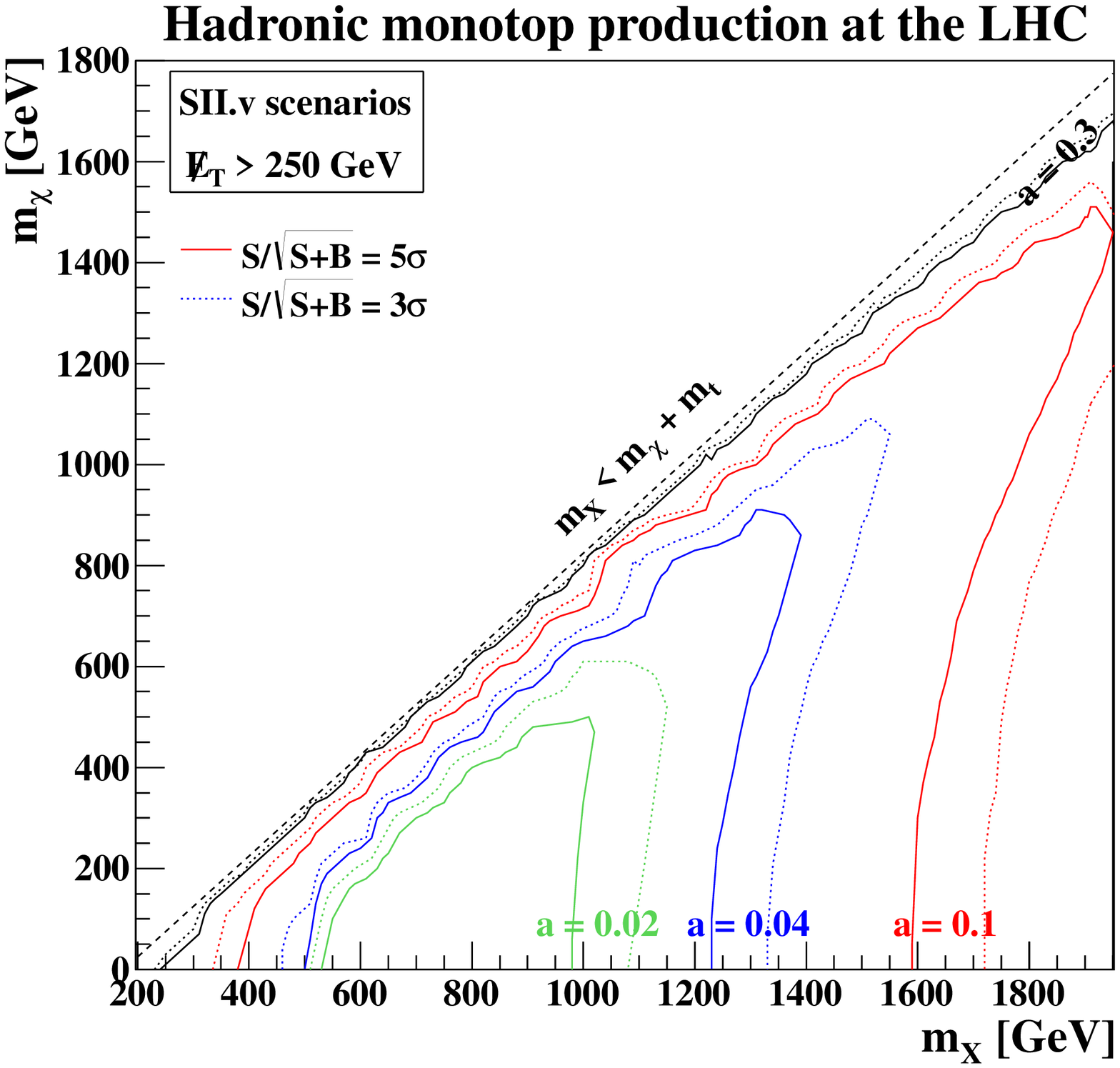}
    \caption{Same as Figure~\ref{fig:monotopsign_b} but
    for scenarios of type {\bf SII}. The results are given in
    $(m_\chi,m_\varphi)$ and $(m_\chi,m_X)$ planes for scenarios of class
    {\bf SII.s} (left panel) and {\bf SII.v} (right panel), respectively, and presented for several values
    of the coupling parameter $a = 0.02$ (green), 0.04 (blue), 0.1 (red) and
    0.3 (black). We focus on the tight $\slashed{E}_T$ requirement which leads to
    better sensitivity in the case of {\bf SII} scenarios and the regions below the plain and dashed
    lines are excluded at the $5\sigma$ and $3\sigma$ levels, respectively.}
    \label{fig:monotopsign_a}
  \end{center}
\end{figure}

Considering a new physics coupling set to $a=0.1$\footnote{Results
for other values of $a$ can be easily deduced as the number of selected
signal events is proportional to $a^2$.} and
20~fb$^{-1}$ of LHC collisions at a center-of-mass energy of 8~TeV,
the number of events surviving all selections are given in Table~\ref{tab:hmonotops},
for the different
background contributions and the series of signal scenarios investigated
so far.
While flavor-changing monotop production (scenarios of class
{\bf SI}) predicts a number of surviving events
depending only on the invisible particle mass (and on the value of the $a$ parameter),
resonant scenarios (of class {\bf SII}) lead to a number of selected events
depending on both the resonant mass, whose monotop production cross section depends on
(together with the coupling parameter),
and on its difference with the sum of the top and the invisible
fermion mass controlling the position of the edge in the missing transverse energy distribution.
The numbers of selected signal ($S$) and background ($B$) events
can be subsequently reexpressed in terms of the LHC sensitivity, that we define
as the significance $s = S/\sqrt{S+B}$, to hadronic monotops. The dependence
of the $s$ quantity on the model parameters is
presented in $(m_\phi,a)$ and $(m_V, a)$ planes for scenarios of class {\bf SI.s}
and {\bf SI.v} on the left and right panels of
Figure~\ref{fig:monotopsign_b}, respectively, and in
$(m_\chi,m_\varphi)$ and $(m_\chi,m_X)$ planes for scenarios
of class {\bf SII.s} and {\bf SII.v} (considering several fixed values of the
$a$ parameter) on the left and right panels of
Figure~\ref{fig:monotopsign_a}. We indicate the contour lines
associated with $s=3$ (dotted curves) and $s=5$ (plain curves)
related to a $3\sigma$ observation
or a $5\sigma$ discovery of a monotop hint at the LHC.

It turns out that the LHC is more sensitive
to scenarios involving vector states, an effect directly related
to the larger associated cross sections compared to the scalar cases.
Next, it is found
that models with larger or moderate couplings of about $a \gtrsim0.1$ are well covered
by both developed search strategies, so that new physics masses ranging up
to 2~TeV are foreseen to be observable at
the LHC in the most optimistic cases. The reaches are reduced in the case of models with a smaller
coupling strength, although the LHC still remains a
promising machine for accessing the low mass regions of the parameter space.
In each case, we present results associated with the $\met$ requirement giving rise
to the best sensitivity, \ie, with the loose criterion ($\met < 150$~GeV) for
flavor-changing monotop production and the tight one ($\met < 250$~GeV) for resonant
production. Generally, the larger missing transverse energy requirement increases,
compared to the looser $\slashed{E}_T$ selection,
the sensitivity to the high
mass parameter space regions whereas it simultaneously decreases
the one to the low mass regions.

Finally, we recall that monotop production induced by a
flavor-changing interaction has been recently searched for by the CDF collaboration
at the Tevatron in the
context
of scenarios where the invisible particle is a new vector state~\cite{Aaltonen:2012ek}.
It has been found that for a coupling strength of
$a=0.1$, benchmark scenarios for which $m_V< 140$~GeV have been
excluded by data. From the results of Figure~\ref{fig:monotopsign_b},
we conclude that future LHC searches for monotops are able to greatly
improve these current constraints.

\subsection{Leptonic monotops at the LHC}\label{sec:leptmonotops}
Hadronic monotops have received special attention in the past as
the associated signature offers the possibility to
use top reconstruction as a tool for background rejection.
In contrast, the leptonic mode
is believed to be more challenging since the branching fraction of the top quark
into a leptonic state is smaller and since there are two different sources of missing
transverse energy, namely a neutrino arising from the top quark leptonic decay
and the new invisible state. However, we choose, in this work, to be
pragmatic and focus on both the hadronic and leptonic channels. First,
the possibility of using leptonic triggers in the context of the LHC experiments
offers a mean to increase the sensitivity to
parameter space regions implying
events with a smaller quantity of missing transverse energy or a smaller cross section.
Next, both channels are
complementary and statistically independent, which offers ways for
further improvements of the obtained limits by combining
the results. Finally, the hadronic
channel is associated with
large systematic uncertainties due, on the one hand, to
the jet energy scale, and, on the other hand, to the
modeling and the resolution of the missing transverse energy. These two sources
of uncertainties are however less critical in the leptonic case, which provides
a further motivation for studying this case.

Event preselection is based on the expected particle content of final states
associated with the signal.
We hence require the presence of exactly one hard and isolated electron
or muon. Its transverse-momentum is demanded to fulfill
$p_T^\ell > 30$~GeV and its pseudorapidity to be compatible with the CMS detector
geometrical acceptance $|\eta^\ell| < 2.5$. Lepton isolation is defined differently
from the choice of Section~\ref{sec:hadrmonotops}
since leptons are used here to define the signal region and not to veto background
events. We compute a variable $I'_{\rm rel}$ where the amount of transverse
energy present in a cone of $R=0.3$, and not $R=0.4$ as in the hadronic
case, is calculated relatively to the lepton
transverse momentum. The significance $s$ of a monotop signature is found
increased when we ask for $I'_{\rm rel} < 0.18$ and $I'_{\rm rel} < 0.06$ for
electrons and muons, respectively, a result largely holding for most of the
reachable parameter space.
In principle, jets or leptons originating from the fragmentation of a heavy
quark can be misidentified as an isolated hard lepton.
However, our detector simulation framework does not account for jets faking
leptons, so that the total `fake' contribution to the background is clearly
not correctly modeled in our analysis. Although we have neglected
this source of background, recent ATLAS and CMS measurements of the single-top
cross section in the $t$-channel production mode have shown that
selection criteria compatible with those introduced below allow us to make
it under good control~\cite{Aad:2012ux,Chatrchyan:2012ep}. Additionally, the
missing transverse energy requirements that are imposed (see below) are tighter than
the one employed in the single top analyses, which further reduces the fake
contamination that is besides driven by QCD multijet subprocesses.

An overwhelming Standard Model background consisting of
about $2\cdot 10^8$ events is expected to pass this selection criterion.
It mainly finds its origin in the production of a leptonically decaying
$W$-boson with jets ($91\%$ of the background) or of a $Z$-boson
with jets ($8\%$ of the background) where one of the produced leptons
is non-reconstructed. Other background processes such as
top-antitop pair, single top and diboson production are all contributing in
a much smaller extent (below 1\%).

In the next selection requirement, we make use of the
configuration of the signal final state which features a $b$-tagged jet.
We then enforce the presence, within the selected events, of
one single $b$-tagged jet lying within the detector acceptance
($|\eta^j| < 2.5$) and with a large transverse momentum $p_T^j > 75$~GeV. This
last criterion has been optimized to allow for a good
background rejection over most of the parameter space and finds its source in the fact that
signal jets can be rather hard as issued from a top quark decay whilst jets
accompanying a singly-produced gauge boson
are in general much softer as arising from
initial state radiation. In addition, events exhibiting the presence
of any other jet with a transverse momentum larger than 20~GeV are vetoed,
so that the (already subdominant)
contamination by single top and top-antitop events is further reduced.
These two selections allow to reduce the number of surviving background events
which are consisting of $W$-boson and $Z$-boson in 88\% and 10\%
of the cases by a factor of more than 700.

\begin{figure}[!t]
  \centering
  \includegraphics[width=.49\textwidth]{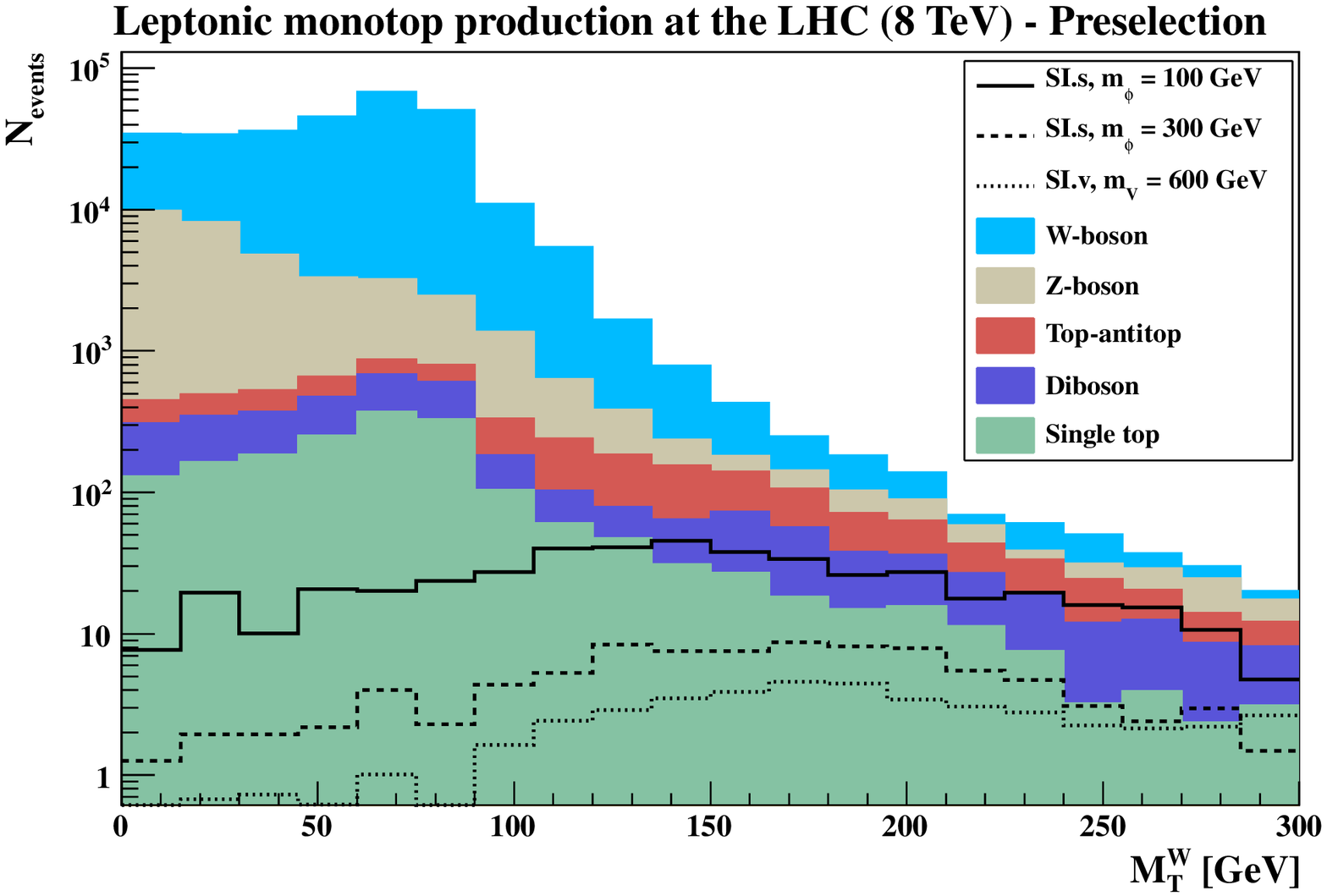}
  \includegraphics[width=.49\textwidth]{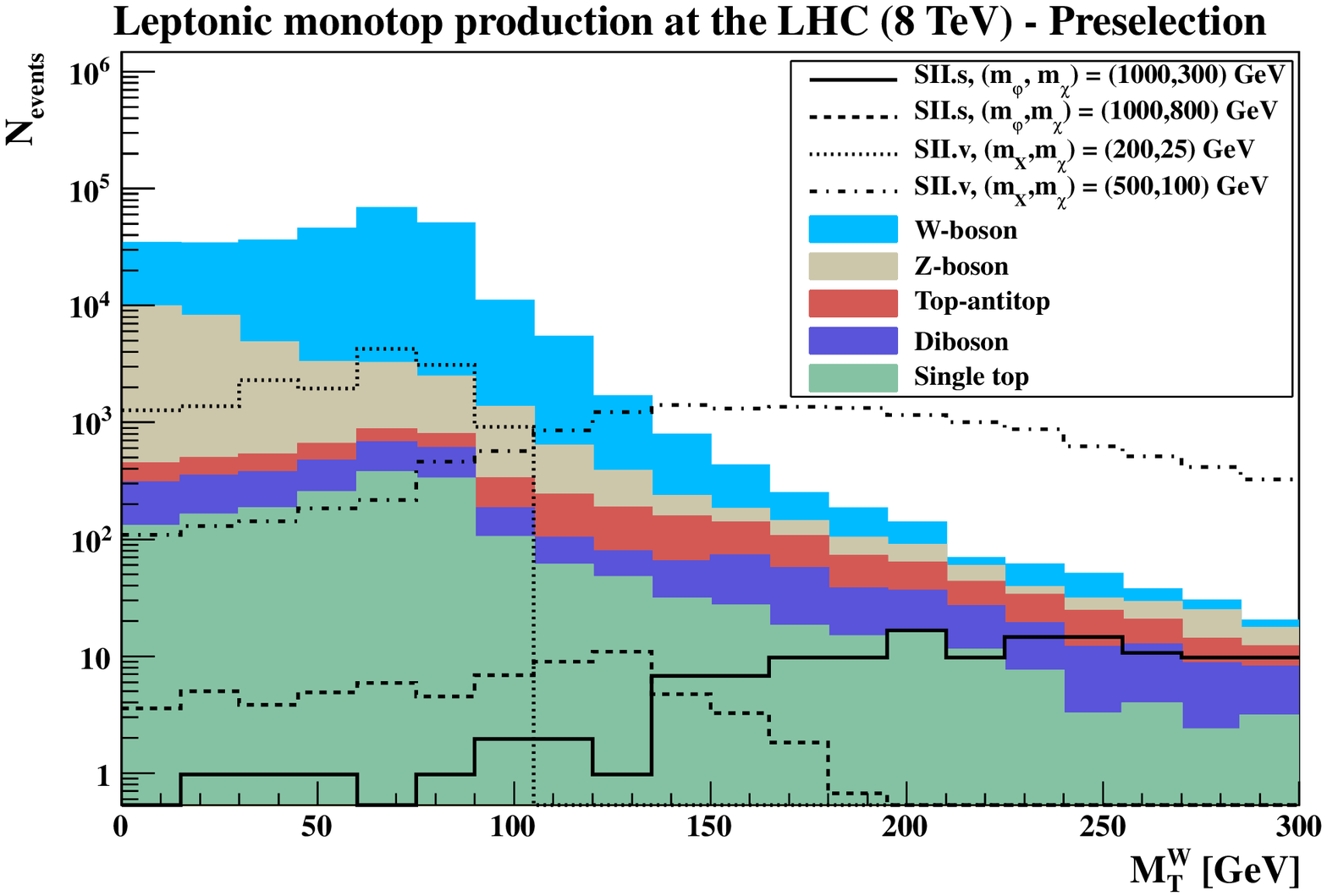}
    \caption{$W$-boson transverse mass spectrum after selecting
    events with one single $b$-tagged jet and one single lepton
    as described in the text. We present separately (and stacked) the various
    contributions to the Standard Model expectation to which we superimpose
    predictions for seven of the eight
    representative signal scenarios of section~\ref{sec:hadrmonotops}.}
  \label{fig:leptonchannel_mtw}
\end{figure}
As stated above, the missing transverse energy present in events
describing the production of a leptonically decaying monotop system
arises from two sources, the neutrino issued from
the decay of the top quark and the invisible new state, in contrast
to Standard Model events where the $\met$ is only originating from
neutrinos or misreconstruction effects.
We benefit from this property
to estimate the $W$-boson transverse mass $M_T^W$ obtained when
considering that all the missing transverse energy finds its origin in a $W$-boson decay,
\be
  M_T^W = \sqrt{2 p_T^\ell \met \Big[1-\cos \Delta \phi_{\ell,\slashed{E}_T} \Big]} \ ,
\ee
where $\Delta\phi_{\ell,
\slashed{E}_T}$ stands for the angular distance, in the azimuthal direction with
respect to the beam, between the lepton and the missing transverse energy. As the
dominant Standard Model contribution is described by events where a real $W$-boson
can be reconstructed, the $M_T^W$ distribution of the background is expected
to peak at smaller $M_T^W$ values, contrary to the signal.
This is illustrated on Figure~\ref{fig:leptonchannel_mtw} where we superimpose
to the background expectation predictions for the eight signal scenarios
introduced in Section~\ref{sec:hadrmonotops}, one of them being already invisible
at that stage. In the case of flavor-changing monotop production,
the signal spectra are very broad and extend to
very large $M_T^W$ values. In contrast, monotop production induced
by the decay of a colored resonance imply distributions that present an edge distorted
by detector effects whose position depends on the resonance and invisible
particle masses. In order to increase the sensitivity to most monotop benchmark
scenarios, we require
\be
  M_T^W > 115~\text{GeV} \ .
\label{eq:mtwreq}\ee
This reduces the background by an extra factor of 40 so that we
expect that about 4000 background events survive, the latter being
constituted of events issued from $W$-boson (77\%), $Z$-boson (9\%), top-antitop (7\%),
diboson (4\%) and single top (3\%) production in association
with jets.
In addition, such a selection on the $W$-boson transverse mass
has also the advantage to render the non-simulated
multijet background under good control~\cite{Aad:2012twa,Chatrchyan:2012bd}.

\begin{figure}[!t]
  \centering
  \includegraphics[width=.49\textwidth]{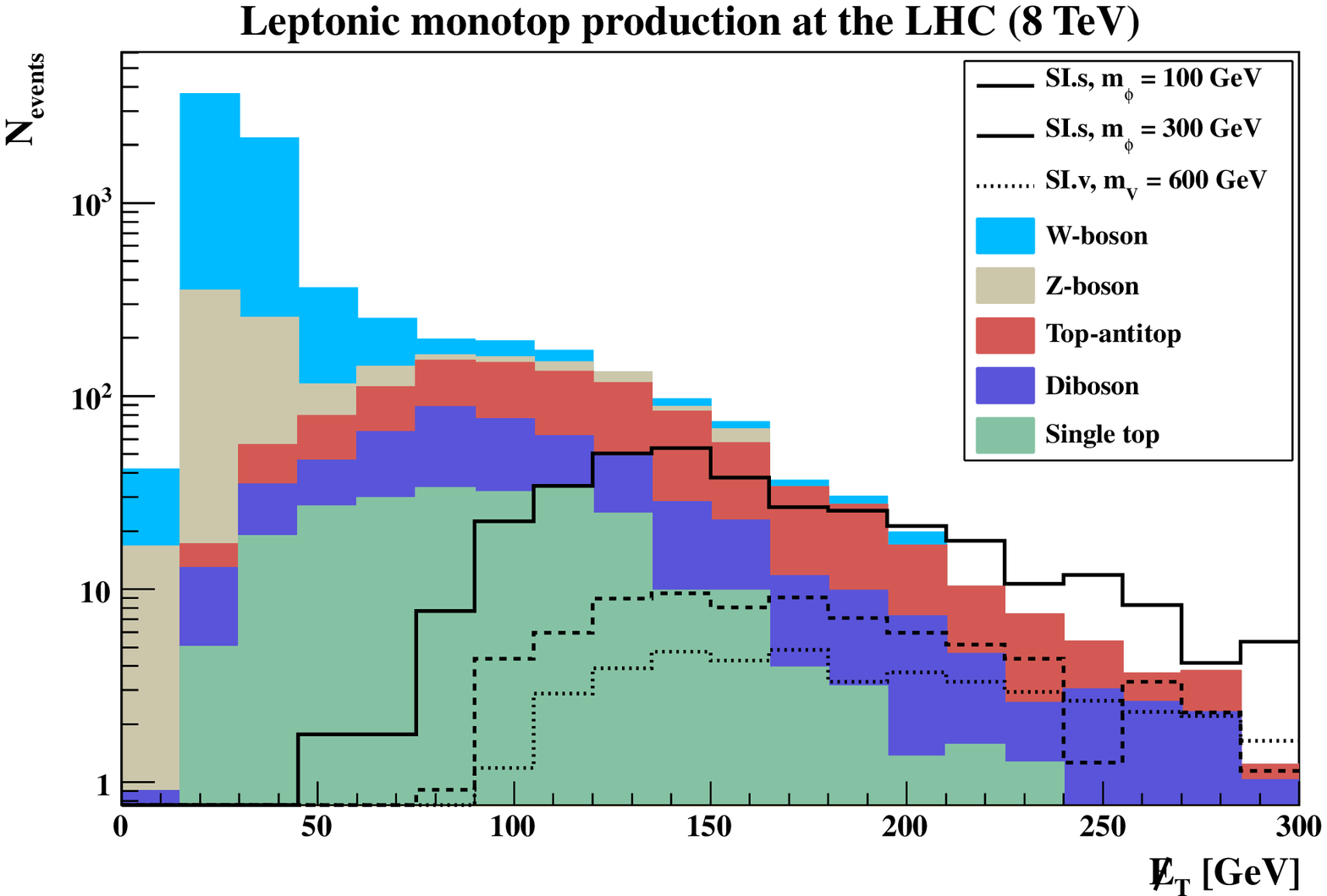}
  \includegraphics[width=.49\textwidth]{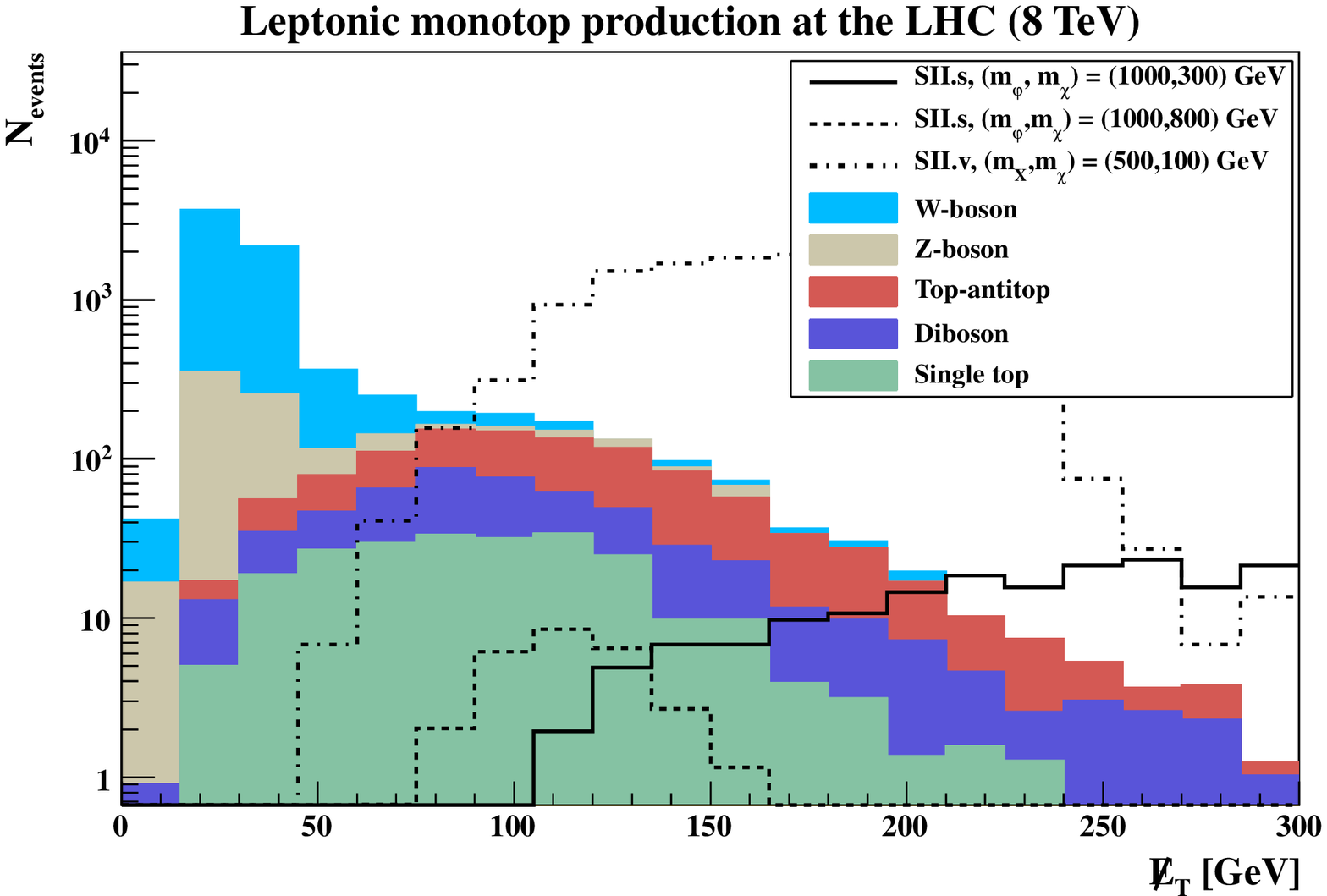}
    \caption{Missing transverse energy distributions after selecting
    events with one single $b$-tagged jet, one single lepton and the requirement
    on the $W$-boson transverse mass described in the text. We present separately
    (and stacked) the various
    contributions to the Standard Model expectation to which we superimpose
    predictions for the eight
    representative signal scenarios of Section~\ref{sec:hadrmonotops}.}
  \label{fig:leptonchannel_met}
\end{figure}

The results of the analysis strategy above are
further depicted on
Figure~\ref{fig:leptonchannel_met} where we present the missing
transverse energy distribution for the different sources of Standard Model
background, together with predictions for the eight monotop benchmark
scenarios introduced in Section~\ref{sec:hadrmonotops}, two of them turning
out to be invisible after the $M_T^W$ selection. While the bulk of the
Standard Model background events lie well within the low missing transverse
energy region, events originating from the considered signal scenarios
are exhibiting peaks whose maximum position lies in general
around larger $\met$ values, their magnitude depending strongly
on the benchmark model. Furthermore, monotop $\met$ spectra also extend
to very large $\met$ values.
In contrast to the hadronic analysis where we can employ
top reconstruction to enhance the signal over background ratio,
no such possibility is available in the leptonic case. Therefore,
in order to get reasonable significance over the entire parameter space,
we choose to optimize the missing transverse energy selection
criterion for each given benchmark scenario so that the quantity
$s$ defined as above, \ie, $s = S/\sqrt{S+B}$, $S$ and $B$ being respectively
the number of selected signal and background events after all the selection
steps, is maximized.

\begin{table}
  \begin{center}
    \begin{tabular}{c||c | c | c | c}
         Process
         & $N_{\rm ev}$, $\met > 120$~GeV
         & $N_{\rm ev}$, $\met > 162$~GeV
         & $N_{\rm ev}$, $\met > 204$~GeV
         & $N_{\rm ev}$, $\met > 225$~GeV\\
       \hline
       \hline
           Top-antitop pair production    & $ 237\pm15 $& $76.94 \pm 8.77$ & $22.70\pm4.76$&$11.77\pm3.43$\\
           $W$-boson production           & $ 27.85\pm5.28 $& $8.35 \pm 2.89$ & $\approx 0$&$\approx 0$\\
           Diboson production             & $ 95.78\pm9.79 $& $40.18 \pm 6.34$ & $18.63\pm4.32$&$13.32\pm3.65$\\
           Single top production          & $ 62.45\pm7.90 $& $14.46 \pm 3.80$ & $5.08\pm2.25$ &$3.13\pm1.77$\\
           $Z$-boson production           & $ 36.71\pm6.06 $& $5.24 \pm 2.29$ & $\approx 0$&$\approx 0$\\
           Other background contributions & $ 0.39\pm0.62 $& $0.29 \pm 0.53$    & $0.19\pm0.43$&$0.10\pm0.31$\\
       \hline
       Total background & $460 \pm 21$ & $145 \pm 12$ & $46.60\pm6.83$ &$28.32\pm5.32$\\
       \hline
       \hline
     {\bf SI.s}, $m_\phi =  100$~GeV  & $300\pm17$ &-&-&- \\
     {\bf SI.s}, $m_\phi =  300$~GeV  &-&$50.41 \pm 7.10$ &-&- \\
     {\bf SI.v}, $m_V =  600$~GeV     &-&-&$28.01\pm5.29$ &- \\
     {\bf SI.v}, $m_V = 1000$~GeV     &-&-&-&$2.64\pm1.62$ \\
    \end{tabular}\vspace{.5cm}
    \begin{tabular}{c||c | c | c | c}
         Process
         & $N_{\rm ev}$, $\met > 192$~GeV
         & $N_{\rm ev}$, $\met > 90$~GeV
         & $N_{\rm ev}$, $\met > 0$~GeV
         & $N_{\rm ev}$, $\met > 90$~GeV\\
       \hline
       \hline
           Top-antitop pair production &$33.85\pm5.82$&$379\pm19$& $536\pm23$&$379\pm 19$\\
           $W$-boson production        &$2.78\pm1.67$&$80.75\pm8.99$& $5720\pm76$&$80.75\pm8.99$\\
           Diboson production          &$23.79\pm4.88$&$174\pm13$& $297\pm17$&$174\pm13$\\
           Single top production       &$7.04\pm2.65$&$129\pm11$& $236\pm15$&$129\pm11$\\
           $Z$-boson production        &$\approx 0$&$62.93\pm7.93$&$687\pm26$&$62.93\pm7.93$\\
           Other background contributions &$0.28\pm0.53$&$0.57\pm0.75$&$0.66\pm0.81$&$0.57\pm0.75$\\
       \hline
       Total background & $67.74 \pm 8.23$&$826\pm29$&$7476\pm86$&$826\pm29$\\
       \hline
       \hline
     {\bf SII.s}, $(m_\varphi,m_\chi) = (1000,300)$~GeV &$269\pm16$&-&-&-\\
     {\bf SII.s}, $(m_\varphi,m_\chi) = (1000,800)$~GeV &-&$26.48\pm5.15$ &-&-  \\
     {\bf SII.v}, $(m_X,m_\chi) = (200,25)$~GeV &-&-& $\approx 0$ &-  \\
     {\bf SII.v}, $(m_X,m_\chi) = (500,100)$~GeV &-&-&-&$12381\pm111$  \\
    \end{tabular}
     \caption{\label{tab:lmonotops}
  Number of events ($N_{\rm ev}$) expected after applying all the selections described in the text,
  for 20 fb$^{-1}$ of LHC collisions at a center-of-mass energy of 8~TeV,
  given together with the associated statistical uncertainties. We present separately the different
  contributions to the Standard Model background and results for the eight
  representative signal scenarios considered in this work. The new physics coupling parameter $a$ is
  set to $0.1$ and the value of the optimized missing transverse energy selection is also indicated.}
  \end{center}
\end{table}

\begin{figure}[t!]
  \centering
  \includegraphics[width=0.49\textwidth]{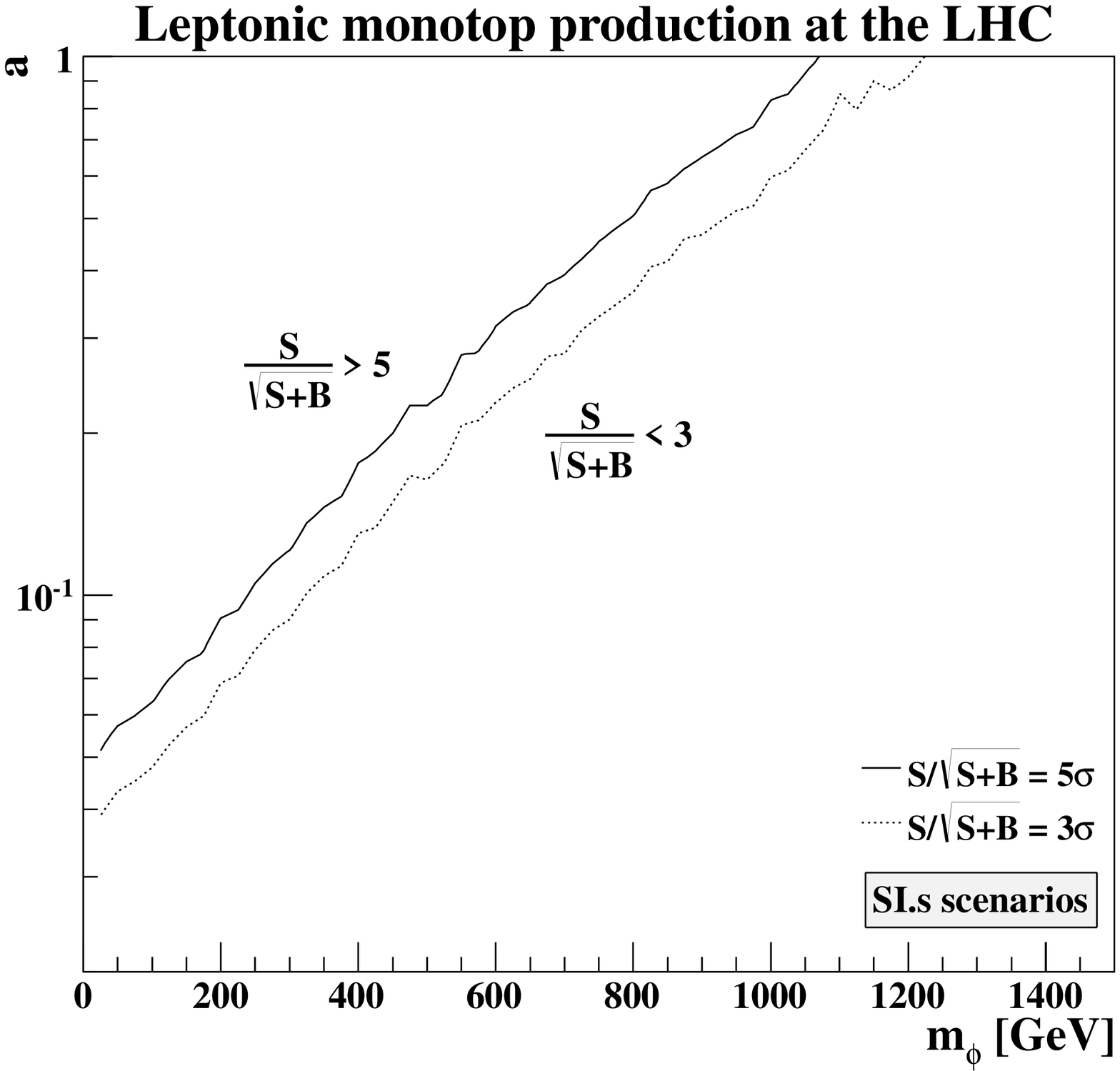}
  \includegraphics[width=0.49\textwidth]{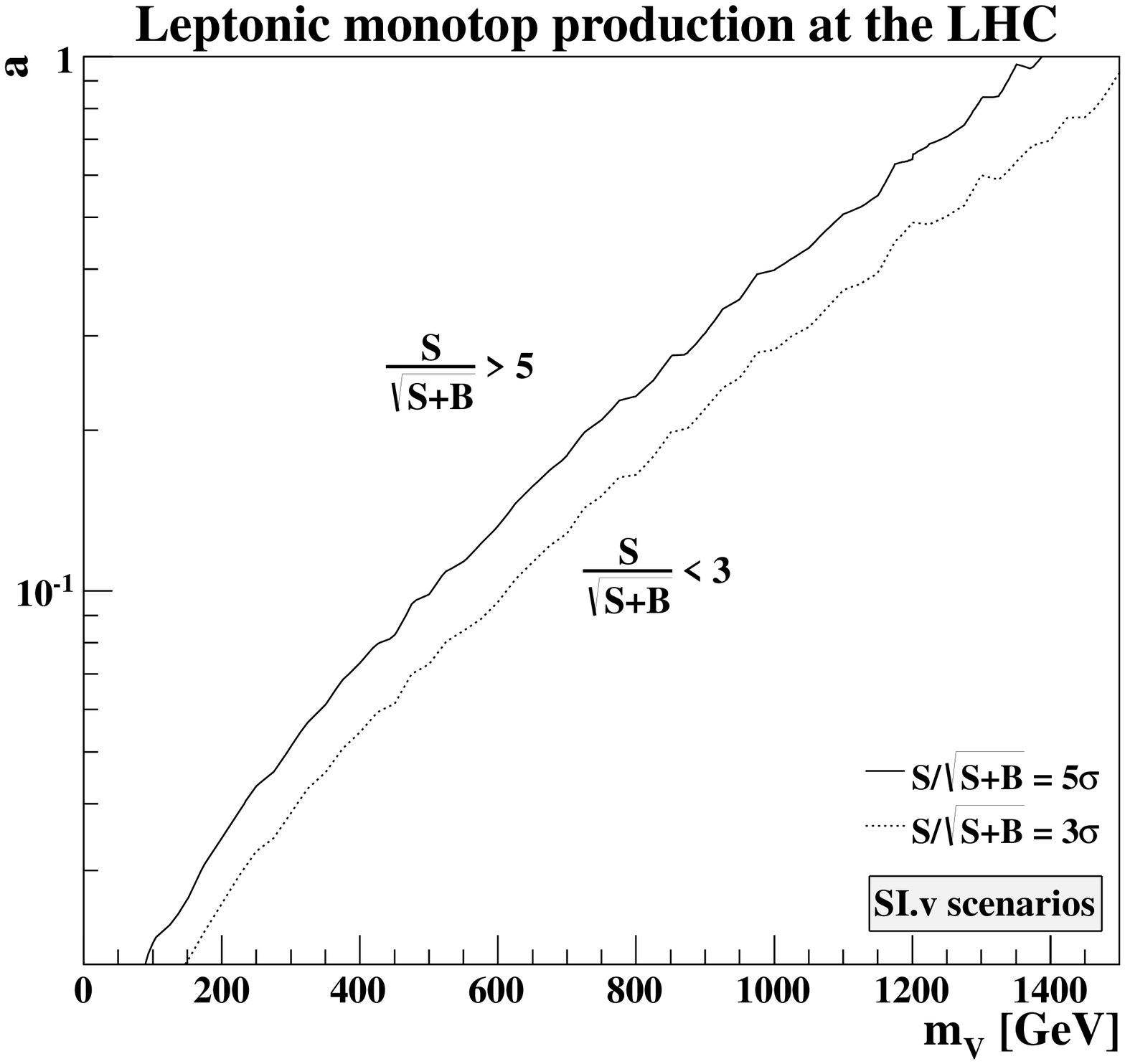}
  \caption{LHC sensitivity to leptonic monotop production in the context of scenarios of
    class {\bf SI} where the Standard Model is supplemented by
    a scalar (left panel) or vector (right panel)
    invisible particle, for 20~fb$^{-1}$ of
    proton-proton collisions at a center-of-mass energy of 8~TeV. The sensitivity
    is calculated as the ratio $S/\sqrt{S+B}$ where $S$ and $B$ are the number of signal
    and background events surviving all selections presented in the text.
    The results are given in $(m_\phi,a)$ and $(m_V,a)$ planes for scenarios of class
    {\bf SI.s} and {\bf SI.v}, respectively.}
    \label{fig:leptonchannel_Exclusion_fcnc}\vspace{.5cm}
  \includegraphics[width=0.49\textwidth]{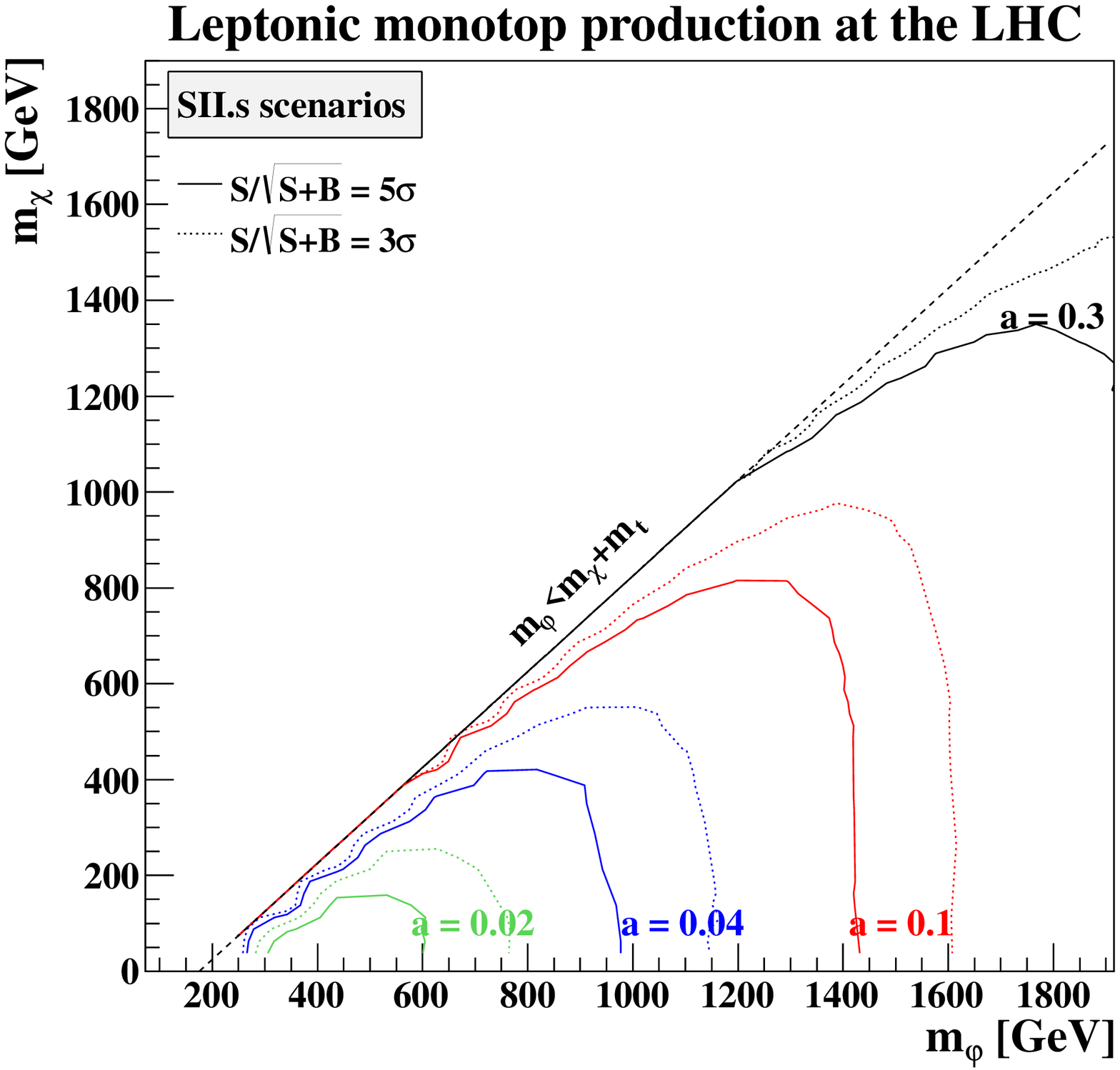}
  \includegraphics[width=0.49\textwidth]{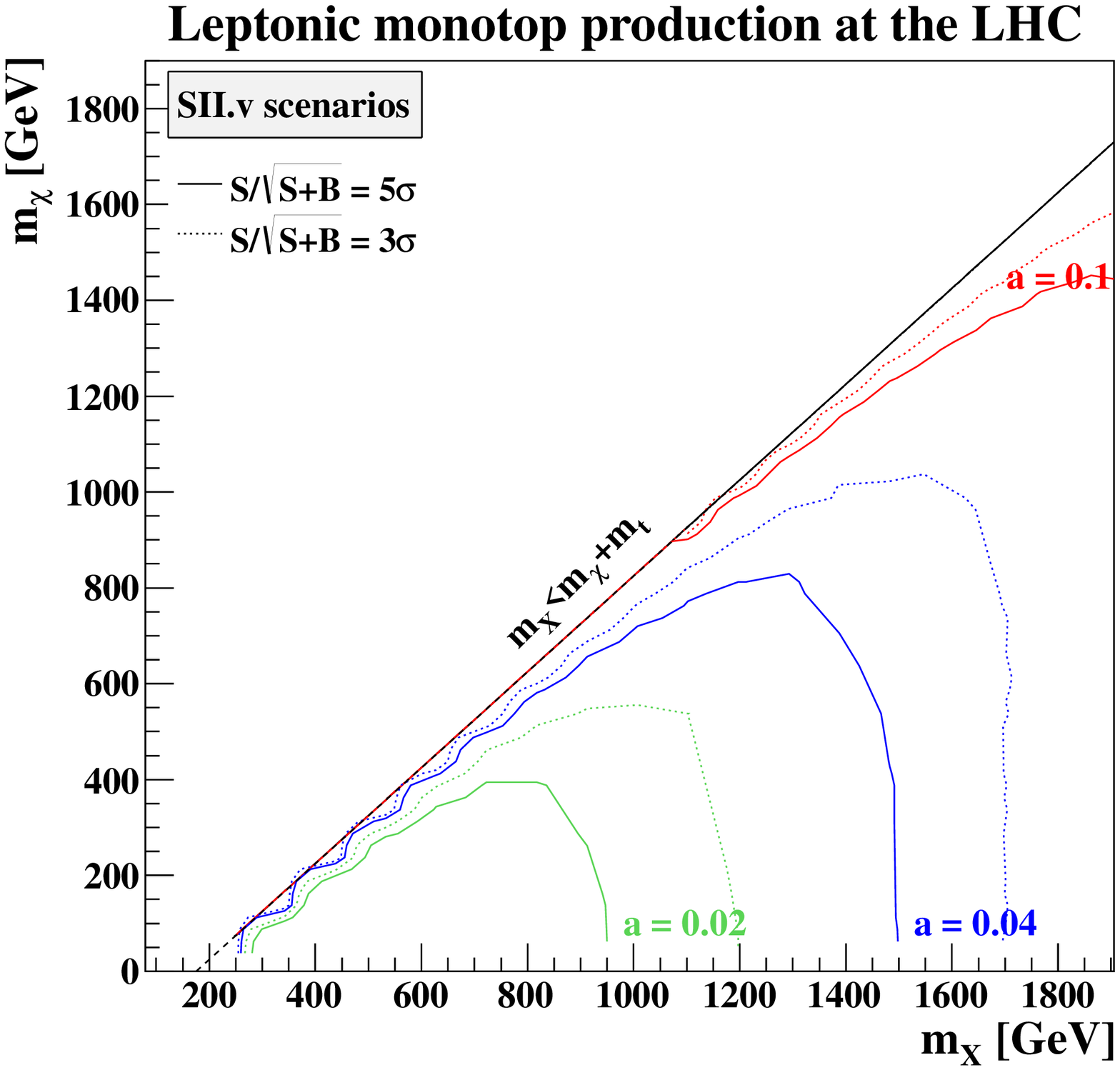}
  \caption{Same as Figure~\ref{fig:leptonchannel_Exclusion_fcnc} but
    for scenarios of type {\bf SII}. The results are given in
    $(m_\chi,m_\varphi)$ and $(m_\chi,m_X)$ planes for scenarios of class
    {\bf SII.s} (left panel) and {\bf SII.v} (right panel),
     respectively, and presented for several values
    of the coupling parameter $a = 0.02$ (green), 0.04 (blue), 0.1 (red) and
    0.3 (black). The regions below the plain and dashed
    lines are excluded at the $5\sigma$ and $3\sigma$ levels, respectively.}
  \label{fig:leptonchannel_Exclusion_res}
\end{figure}

We present on Table~\ref{tab:lmonotops} the number of remaining events after
all selections
for the different contributions to the Standard Model background as well as
for the eight monotop scenarios
of Section~\ref{sec:hadrmonotops}. Our predictions are normalized to
20~fb$^{-1}$ of LHC collisions at a center-of-mass
energy of 8~TeV and in each case, we include
the derived value for the optimized missing transverse energy requirement.
When the monotop state is produced via a flavor-changing
neutral interaction and for a new physics coupling strength of $a=0.1$,
the number of expected signal events is in general relatively large when compared to the background
expectation, at least for invisible masses below 600~GeV.
Consequently, we expect that an important
fraction of the monotop parameter space for scenarios of class {\bf SI} can be reachable
with the available
integrated luminosity for collisions at $\sqrt{S_h}=8$~TeV. This
is illustrated on Figure~\ref{fig:leptonchannel_Exclusion_fcnc} where we present,
in $(m_\phi,a)$ (left panel of the figure) and $(m_V, a)$ (right panel
of the figure) planes contours associated with a $3\sigma$ and $5\sigma$
significance for scenarios of class {\bf SI.s} and {\bf SI.v}, respectively. The
results are found to be comparable to the hadronic case, although
the sensitivity to the larger mass regions of the
parameter space is slightly increased.
In the second part of the table and on
Figure~\ref{fig:leptonchannel_Exclusion_res}, we turn to the investigation
of the LHC sensitivity to the resonant production of  leptonically decaying monotops.
We again obtain conclusions similar to the hadronic case with a slight increase in sensitivity
for the high mass region of the parameter space, as shown in the
$(m_\chi,m_\varphi)$ and $(m_\chi,m_X)$ planes of scenarios
of class {\bf SII.s} and {\bf SII.v} (considering several fixed values of the
$a$ parameter) on the left and right panels of
Figure~\ref{fig:leptonchannel_Exclusion_res}.


\section{Conclusions}
\label{sec:conclusions}
In this paper, we have investigated in details the LHC phenomenology
of the monotop signature. We have considered a set of representative scenarios
based on a simplified model description that
feature all the possible production modes of a monotop state at a
hadronic collider. It has been shown that,
by means of state-of-the-art Monte Carlo simulations including a modeling
of a CMS-like detector response of the 20~fb$^{-1}$
of collisions that
have been produced at the 2012 LHC run at a center-of-mass energy
of 8~TeV, a large fraction of the parameter space is in principle reachable.
In our work, we have followed a pragmatic approach and
studied both the hadronic and leptonic decay modes of the monotop system,
even though the last case might be naively thought as more challenging to observe.
We have designed several dedicated analyses allowing to derive the new physics
masses and couplings that are observable at the $3\sigma$ and $5\sigma$ level. In
particular and for couplings of about $a=0.1$,
masses ranging up to 300~GeV (500~GeV) could be observed when the monotop
state is produced via flavor-changing neutral interactions with a new scalar
(vector) state. In contrast, monotop resonant production allows to access beyond
the Standard Model masses ranging up to about 1~TeV (1.5~TeV) when the Standard Model contains
a new scalar (vector) colored field. Those results are very encouraging and should
motivate further investigations by both the ATLAS and CMS experiments
in the context of real data.

\acknowledgments
The authors are grateful to K.~Kotov for lively discussions on the topic.
This work has been partially supported by the Theory-LHC France-initiative of
the CNRS/IN2P3, by the French ANR 12 JS05 002 01 BATS@LHC and by
a Ph.D.\ fellowship of the French ministry for education and research.


\bibliography{biblio}

\end{document}